\documentclass[prd,amsmath,aps,floats,amssymb,floatfix,superscriptaddress,nofootinbib,twocolumn,preprintnumbers]{revtex4-2}
\UseRawInputEncoding
\usepackage{tabularx}
\usepackage{amsmath,amssymb,natbib,latexsym,times}
\usepackage[T1]{fontenc}
\usepackage{aecompl}

\usepackage{multirow}
\usepackage{rotating}
\usepackage{booktabs}
\usepackage{soul}
\usepackage{enumitem}
\usepackage{verbatim}
\usepackage[usenames,dvipsnames]{xcolor}
\usepackage{tabulary}
\usepackage{tabularx}
\usepackage{todonotes}
\usepackage{hyperref}
\hypersetup{colorlinks=true, linkcolor=blue, citecolor=blue, filecolor=blue, urlcolor=blue}
\usepackage{ulem}

\let\oldequation\equation
\let\oldendequation\endequation
\renewenvironment{equation}
  {\linenomathNonumbers\oldequation}
  {\oldendequation\endlinenomath}

\usepackage{hyperref}
\usepackage[T1]{fontenc}
\usepackage{ae,aecompl}
\usepackage{graphicx}	% Including figure files
\usepackage{amsmath}	% Advanced maths commands
\usepackage{adjustbox}
\usepackage{multirow}
\usepackage{newtxmath}  % For Indicator function. See https://tex.stackexchange.com/questions/26637/, answer by xovee.
\usepackage{booktabs}
\definecolor{purple}{RGB}{150,0,200}

\usepackage{lineno}
\newcommand{\maglim}{\textsc{MagLim}\xspace}
\newcommand{\maglimpp}{\textsc{MagLim++}\xspace}

\newcommand{\mdet}{\textsc{METADETECT}\xspace}

\usepackage{xcolor}
\definecolor{Gray}{gray}{0.9}
\definecolor{myorange}{RGB}{255, 104, 51}

\usepackage{ulem}

\usepackage{xspace} %This allows to add extra space when needed in new commands

\defcitealias{y6-gold}{\textsc{Y6Gold}}
\defcitealias{y6-piff}{\textsc{Y6PSF}}
\defcitealias{y6-mask}{\textsc{Y6Mask}}
\defcitealias{y6-balrog}{\textsc{Y6Balrog}}
\defcitealias{y6-metadetect}{\textsc{Y6Shear}}
\defcitealias{y6-maglim}{\textsc{Y6Pos}}
\defcitealias{y6-sourcepz}{\textsc{Y6ShearPZ}}
\defcitealias{y6-lenspz}{\textsc{Y6PosPZ}}
\defcitealias{y6-wz}{\textsc{Y6WZ}}
\defcitealias{y6-nzmodes}{\textsc{Y6Modes}}
\defcitealias{y6-magnification}{\textsc{Y6Mag}}
\defcitealias{y6-gglens}{\textsc{Y6GGL}}
\defcitealias{y6-methods}{\textsc{Y6Model}}
\defcitealias{y6-ppd}{\textsc{Y6PPD}}
\defcitealias{y6-imagesims}{\textsc{Y6Imsim}}
\defcitealias{y6-cardinal}{\textsc{Y6Mock}}
\defcitealias{y6-1x2pt}{\textsc{Y6-1x2pt}}
\defcitealias{y6-2x2pt}{\textsc{Y6-2x2pt}}
\defcitealias{y6-3x2pt}{\textsc{Y6-3x2pt}}
\defcitealias{y6-extensions}{\textsc{Y6Extend}}

\begin{document}
\title[DES Y6 Galaxy-galaxy lensing]{Dark Energy Survey Year 6 Results: \\ Galaxy-galaxy lensing}
% Author list file generated with: mkauthlist 1.4.0 
% mkauthlist -f -s --aux ggl_first_tier.csv ggl_authors.csv ggl_authorlist.tex 

\author{G.~Giannini}\email{giulia.giannini15@gmail.com}
\affiliation{Institute of Space Sciences (ICE, CSIC),  Campus UAB, Carrer de Can Magrans, s/n,  08193 Barcelona, Spain}
\affiliation{Department of Astronomy and Astrophysics, University of Chicago, Chicago, IL 60637, USA}

\author{G.~Camacho-Ciurana}
\affiliation{Institute of Space Sciences (ICE, CSIC),  Campus UAB, Carrer de Can Magrans, s/n,  08193 Barcelona, Spain}

\author{A.~Whyley}
\affiliation{Institute of Cosmology and Gravitation, University of Portsmouth, Portsmouth, PO1 3FX, UK}

\author{J.~Prat}
\affiliation{Nordita, KTH Royal Institute of Technology and Stockholm University, Hannes Alfv\'ens v\"ag 12, SE-10691 Stockholm, Sweden}
\affiliation{DARK, Niels Bohr Institute, University of Copenhagen, Blegdamsvej 17, 2100, Copenhagen, Denmark}

\author{J.~Blazek}
\affiliation{Department of Physics, Northeastern University, Boston, MA 02115, USA}

\author{C.~S{\'a}nchez}
\affiliation{Departament de F\'{\i}sica, Universitat Aut\`{o}noma de Barcelona (UAB), 08193 Bellaterra, Barcelona, Spain}
\affiliation{Institut de F\'{\i}sica d'Altes Energies (IFAE), The Barcelona Institute of Science and Technology, Campus UAB, 08193 Bellaterra (Barcelona) Spain}

\author{G.~Zacharegkas}
\affiliation{Kavli Institute for Cosmological Physics, University of Chicago, Chicago, IL 60637, USA}

\author{A.~Alarcon}
\affiliation{Institute of Space Sciences (ICE, CSIC),  Campus UAB, Carrer de Can Magrans, s/n,  08193 Barcelona, Spain}

\author{E.~Legnani}
\affiliation{Institut de F\'{\i}sica d'Altes Energies (IFAE), The Barcelona Institute of Science and Technology, Campus UAB, 08193 Bellaterra (Barcelona) Spain}

\author{A.~Amon}
\affiliation{Department of Astrophysical Sciences, Princeton University, Peyton Hall, Princeton, NJ 08544, USA}

% ...

\collaboration{DES Collaboration}

\label{firstpage}

\begin{abstract}
We present galaxy--galaxy lensing (GGL) measurements from the full six years of data from the Dark Energy Survey (DES Y6), covering $4031\,\mathrm{deg}^2$ and used in the DES Y6 $3\times2$pt cosmological analysis. We use the \textsc{MagLim++} lens sample, containing $\sim 9$ million galaxies divided into six redshift bins, and the \textsc{Metadetection} source catalog, including $\sim 140$ million galaxies divided into four redshift bins. The mean tangential shear signal achieves a total signal-to-noise ratio (S/N) of $173$, corresponding to a $17\%$ improvement over DES Y3. After applying the scale cuts used in the cosmological analysis, with $R_{\min}=6\,\mathrm{Mpc}/h$ ($4\,\mathrm{Mpc}/h$) for the linear (nonlinear) galaxy-bias model, the S/N is reduced to $75$ (90). A comprehensive suite of validation tests demonstrates that the measurement is robust against observational and astrophysical systematics at the statistical precision required for the DES Y6 analysis. Although not used in the main cosmological analysis, we extract high--signal-to-noise geometric shear-ratio measurements from the galaxy--galaxy lensing signal on small angular scales. These measurements provide an internal consistency check on the photometric redshift distributions and shear calibration used in the $3\times2$pt analysis.
\end{abstract}

\vspace{0.2cm}

\preprint{DES-2025-0955}
\preprint{FERMILAB-PUB-26-0031-PPD}
\maketitle
\setcounter{footnote}{1}
%%%%%%%%%%%%%%%%%%%%%%%%%%%%%%%%%%%%%%%%%%%%%%%%%%

%%%%%%%%%%%%%%%%% BODY OF PAPER %%%%%%%%%%%%%%%%%%

%_____________________INTRODUCTION_________________________
\section{Introduction}
\label{sec:intro}

The phenomenon of gravitational lensing, predicted by Einstein's theory of General Relativity, occurs when light propagating through curved spacetime is deflected in the presence of mass. In cosmology, weak gravitational lensing by large-scale structure arises from the cumulative effect of matter along the line of sight, producing small, coherent distortions in the observed shapes of distant galaxies that can be used as a powerful probe of the matter distribution and cosmological parameters

Galaxy-galaxy lensing (GGL) measures the weak gravitational lensing signal produced when matter distributions surrounding foreground (lens) galaxies distort the shapes of background (source) galaxies. This manifests as tangential shape distortions of source galaxies around lens galaxies, which we capture by averaging over many lens-source pairs to estimate the mean tangential shear. This quantity is a two-point correlation function that quantifies the statistical relationship between lens galaxy positions and source galaxy shape distortions. Two-point statistics excel in large-scale regimes where density fluctuations remain predominantly Gaussian, efficiently capturing most cosmological information while offering analytical tractability and well-understood systematics. 

The mean tangential shear is sensitive to both the geometrical properties of the Universe and the growth of structure through matter fluctuations. However, since we trace the matter distribution using galaxies, GGL measurements depend on the galaxy bias factor--the relationship between galaxy and matter distributions--which introduces degeneracies with cosmological parameters like $\sigma_8$ (the amplitude of matter fluctuations). To break these degeneracies and extract robust cosmological information, GGL is typically combined with complementary probes.

This paper is part of the final Dark Energy Survey Year 6 (DES Y6) cosmological analysis, using the complete six-year observational dataset. In particular, it is part of the powerful ``3$\times$2pt'' methodology, which combines three two-point correlation measurements: galaxy-galaxy lensing (position-shear), galaxy clustering (position-position), and cosmic shear (shear-shear), see \cite{PratBacon2025} for an accessible introduction. This combination has proven remarkably effective at breaking parameter degeneracies and also at being robust against systematic uncertainties, due to its ability to self-calibrate them. Within the 3$\times$2pt framework, besides providing important additional cosmological information, GGL plays a crucial calibration role for multiple components: (i) galaxy bias parameters, through its different dependency than galaxy clustering; (ii) intrinsic alignment parameters, particularly through measurements in overlapping lens-source bins; (iii) lens magnification parameters; and (iv) both lens and source redshift uncertainty parameters through the inherent ``shear-ratio''  information.

The shear-ratio technique compares tangential shear measurements from the same lens sample but different source redshift bins. In such ratios, lens-dependent properties (galaxy bias, matter power spectrum) cancel out, yielding a purely geometric measurement when the thin-lens approximation applies. This geometric sensitivity provides constraints on source and lens redshift distributions, and shear calibration parameters. Moreover, because shear-ratio modeling depends only on geometry, we can use such measurements to probe arbitrarily small scales--beyond the limits where galaxy bias and baryonic feedback otherwise compromise our modeling of the nonlinear regime. The DES collaboration has used such ratios in different ways in each data realease: in Y1, using small-scale shear-ratio measurements to test redshift calibration; in Y3, incorporating them directly into the 3$\times$2pt analysis to extend the usable range of tangential shear measurements \citep{y3-shearratio}. For Y6, we return to purely geometric ratios as an independent test of redshift calibration, which we present in this work.

\begin{table}
    \centering
    \caption{Reference table for supporting papers to this work.} 
    \label{tab:alias}
        \begin{tabular}{lp{1.5cm}p{5cm}}
            \hline 
            \hline
            Shorthand  & Reference & Description \\ 
        \hline
             \textsc{Y6Gold} &   \cite{y6-gold}   & Coadd object catalog and associated products \\

             \textsc{Y6PSF} &   \cite{y6-piff}   & Point spread function (PSF) models  \\

             \textsc{Y6Shear} &   \cite{y6-metadetect}   & \textsc{MetaDetect} shear catalog  \\
             \textsc{Y6Pos} &   \cite{y6-maglim}   & \textsc{MagLim++} lens selection and galaxy clustering \\
             \textsc{Y6Balrog} &   \cite{y6-balrog}   & Synthetic source injection  \\
            \textsc{Y6Mask} &   \cite{y6-mask}   & Survey mask  \\
             \textsc{Y6ShearPZ} &   \cite{y6-sourcepz}   & Source photo-$z$ calibration  \\
             \textsc{Y6PosPZ} &   \cite{y6-lenspz}   & Lens photo-$z$ calibration  \\
             \textsc{Y6WZ} &   \cite{y6-wz}   & Clustering redshift  \\
             \textsc{Y6Mode} &   \cite{y6-nzmodes}   & Method for marginalizing redshift   \\
             &  &  uncertainty \\
             \textsc{Y6Imsim} &   \cite{y6-imagesims}   & Image simulation  \\
             \textsc{Y6Mock} &   \cite{y6-cardinal}   & Mock galaxy catalog  \\
             \textsc{Y6Model} &   \cite{y6-methods}   & Modeling and analysis choices  \\
             \textsc{Y6Mag} &   \cite{y6-magnification}   & Characterising lens magnification  \\
             \textsc{Y6PPD} &   \cite{y6-ppd}   & Quantifying internal consistency with posterior predictive distribution \\
             \textsc{Y6GGL} &   \textit{this work}   & Galaxy-galaxy lensing \\
             \textsc{Y6-1x2pt} &   \cite{y6-1x2pt}   & Cosmology from cosmic shear  \\
             \textsc{Y6-2x2pt} &   \cite{y6-2x2pt}   & Cosmology from galaxy-galaxy lensing and galaxy clustering \\
          \textsc{Y6-3x2pt} &  \cite{y6-3x2pt}   & Cosmology from cosmic shear, galaxy-galaxy lensing and galaxy clustering \\
             \hline
             \hline
        \end{tabular}
\end{table}

For the DES Y6 3$\times$2pt analysis, we continue to use a magnitude-limited sample for our tracer or lens galaxies, building upon the approach established in DES Y3. For the DES Y6 analysis, we start with the baseline Y3 \textsc{MagLim} selection but then apply a series of novel quality cuts to identify and remove contamination in the sample resulting in what we call \textsc{MagLim++} \citepalias{y6-maglim}. Because of the similar selection, the \textsc{MagLim++} lens sample maintains a number density similar to Y3.
%These galaxies typically reside in dark matter halos with masses of $\log_{10}(M_h/M_\odot) \simeq 13.5$ \citealt{Zacharegkas2022}.
For our source catalog, presented in \citetalias{y6-metadetect}, we employ the new \textsc{Metadetection} methodology---an important evolution from the \textsc{Metacalibration} technique used in previous analyses. While \textsc{Metacalibration} \citep{Sheldon2023} applies shears to already-detected objects, \textsc{Metadetection} applies shears to the original images before object detection occurs. This advancement allows us to calibrate biases introduced by the object detection process itself, which can be particularly significant for faint galaxies near the detection threshold. The resulting Y6 shape catalog contains approximately 1.5 times more galaxies than in DES Y3. This increase is primarily driven by the greater survey depth in Y6--owing to roughly twice the number of exposures over a comparable footprint--rather than by changes in the shear-calibration methodology.

The photometric redshift calibration has also improved: we included broad band $g$, because of improvements in the color-dependent PSF modeling and star-galaxy separation \citepalias{y6-piff}. We introduced a more comprehensive framework to estimate redshift sample uncertainties, capturing residual biases and catalog incompleteness; a new approach based on Bernstein et al. (2025) is taken to parametrize the uncertainty in the redshift distributions that allows for more flexible and general variations in the distributions, rather than relying on simple shift-and-stretch parametrizations.

Overall, the DES collaboration has achieved substantial improvements in the precision of GGL measurements across its different data releases. In DES Y1 \citep{y1-gglensing}, we measured tangential shear with a signal-to-noise ratio (S/N) of 73 for the full range of scales (from 2.5 to 250 arcmin) and redshift bins. For DES Y3 \citep{y3-gglensing}, while maintaining similar depth but quadrupling the survey area, the S/N increased  to 148. The current DES Y6 dataset maintains approximately the same area as Y3 but doubles the depth, with our source sample containing $\sim$1.5 times more galaxies than Y3, resulting in S/N of 173. Notably, throughout the years of DES analysis, the tangential shear measurement has consistently proven to be remarkably robust against observational uncertainties. It has been found to be effectively free of systematics in all previous DES data releases. Even when extending the analysis to a full decade of additional scales on the small-scale end (down to 0.25 arcmin from the fiducial limit of 2.5 arcmin), measurements remained free of observational systematic effects \citep{Zacharegkas2022}. This robustness can be partially attributed to the nature of GGL as a cross-correlation between lens galaxy positions and source galaxy shears, making it inherently less susceptible to systematic effects that might contaminate either catalog individually. 

From a modeling perspective, however, GGL faces significant challenges, particularly in the treatment of small-scale galaxy bias, which exhibits highly nonlinear and stochastic behavior. Additional uncertainties arise from the modeling of the nonlinear matter power spectra. Our strategies for addressing these modeling limitations have evolved across survey generations. In DES Y1, we discarded data points with projected lens-source separations below 12 Mpc/$h$. In DES Y3, despite the increased S/N, we were able to incorporate scales down to 6 Mpc/$h$ by implementing point-mass marginalization following the methodology of \cite{MacCrann2020}. Though this approach significantly reduces the effective S/N (in DES Y3, from 148 with full scales to 67 after applying 6 Mpc/$h$ scale cuts, and further to 32 with point-mass marginalization), it has been demonstrated that it provides greater constraining power than implementing more aggressive scale cuts \citep{Prat2023}. For DES Y6, even with the significant increase in measurement precision, we retain the physically motivated scale cuts used in previous analyses--6 Mpc/$h$ for the linear bias model and 4 Mpc/$h$ when allowing for nonlinear galaxy bias, while also improving the modeling of the matter power spectrum. With these choices, the total signal-to-noise ratio is 75 (90) for the linear (nonlinear) bias model, and decreases to 39 (48) when point-mass marginalisation is accounted for. Specifically, we have transitioned from the revised \textsc{Halofit} prescription of \cite{Takahashi2012} used in DES Y3, which provides nonlinear matter power spectra fit to $N$-body simulations, to the more sophisticated \textsc{HMCODE} from \citep{Mead2021}, which incorporates prescriptions for nonlinear cosmological power spectra with baryonic feedback effects.

This series of papers marks the culmination of the DES and will serve as the legacy galaxy-galaxy lensing measurements from the Dark Energy Survey. 

The paper is organized as follows. In Section 2 we describe the data sets used in this analysis. Section 3 presents the theoretical modeling adopted throughout the paper, while Section 4 details the measurement methodology, presenting the tangential shear estimator. In Section 5 we assess the robustness of our results through a series of systematic tests. Section 6 focuses on the shear-ratio analysis, and Section 7 summarizes our findings and presents the main conclusions.

To ease the reader's navigation, this paper refers to all the DES Y6 $3\times2pt$ supporting papers with the labels listed in Table~\ref{tab:alias}.

\section{Data}
\label{sec:data}

We use the full six years of observations by the Dark Energy Survey (DES) in support of static-sky cosmology analyses.

\begin{figure}
    \centering
    \includegraphics[width=0.45\textwidth]{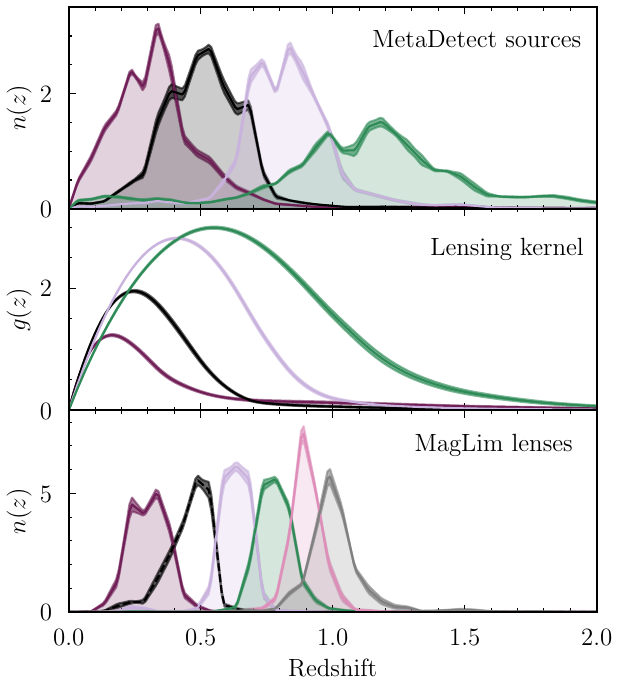}
    \caption{Redshift distributions for the source (top) and lens (bottom) samples, with the corresponding lensing efficiency shown in the middle panel. In all panels, solid lines indicate the mean, while shaded regions denote the standard deviation across 100 realizations of reconstructed $n(z)$, sampled from the posterior of our fiducial $3\times2$pt linear bias analysis in which lens bin 2 is excluded. For lens bin 2 (shown with unfilled lines), the $n(z)$ realizations are instead drawn from a $3\times2$pt linear bias analysis including all bins.}
    \label{fig:nz}
\end{figure}

\begin{table*}
    \centering
    \caption{\label{tab:shapes} Characteristics of the source and lens samples. For each tomographic bin and for all objects, we list the number of objects, effective number density, shape noise, shear response and weighted residual mean shear for the source sample, and the number of objects and number density for the lens sample
    }
     \begin{tabular}{c|cccccc|cc}
\hline
        & \multicolumn{6}{c}{Source sample} & \multicolumn{2}{c}{Lens sample}  \\
     Bin & num. & $n_{\rm eff}$  & $\sigma_{\rm e}$ & $\langle R \rangle$ & $\langle e_1 \rangle $ & $\langle e_2 \rangle $ & num. & $n_{\rm gal}$\\ 
        & of object &  [gals/arcmin$^2$] & && $ \times 10^{-4}$ & $ \times 10^{-4}$ & of object & [gals/arcmin$^2$]\\     
     \hline 
      % \vspace{0.1cm}
        1 & 33,707,071 & 2.05 & 0.265 & 0.856 & +1.6 & +0.62 & 1,852,538 & 0.128 \\
        2 & 34,580,888 & 2.10 & 0.287 & 0.869 & +0.45 & -0.58 & 1,335,294 & 0.092 \\
        3 & 34,646,416 & 2.14 & 0.282 & 0.819 & +2.4 & -1.7 & 1,413,738 & 0.097 \\
        4 & 36,727,798 & 2.32 & 0.347 & 0.692 & +3.1 & +1.1 & 1,783,834 & 0.123 \\
        5 & - & - & - & - & - & - & 1,391,521 & 0.096 \\
        6 & - & - & - & - & - & - & 1,409,280 & 0.097 \\
        All & 139,662,173 & 8.29 & 0.289 & 0.819 & +1.4 & -0.15 & 9,186,205 & 0.633 \\
\hline
    \end{tabular}\label{tab:data}
\end{table*}

\subsection{Lens sample: \textsc{MagLim++}}

The \maglimpp Y6 sample is a magnitude limited sample selected from the DES Y6 Gold catalog. DES Y6 Gold \citepalias{y6-gold} is a curated data set derived from DES Data Release 2 (DR2) that incorporates improved measurement, photometric calibration, object classification and value added information. Y6 Gold comprises nearly 5000 deg$^2$ of $grizY$ imaging in the south Galactic cap and includes 669 million objects with a depth of $i_{AB} \sim 23.4$ mag at S/N $\sim 10$ for extended objects. After quality selections, benchmark samples contain 448 million galaxies and 120 million stars.

The primary \maglimpp sample selection builds upon the DES~Y3 \maglim definition \citep{y3-2x2maglimforecast}, starting from galaxies in the \textsc{Gold} catalog. A linear relation between $i$-band magnitude and the \textsc{DNF} photometric redshift \citep{DeVicente2016} is used to define the selection, effectively removing faint, high-redshift galaxies while preserving a roughly constant signal-to-noise ratio across redshift:
\begin{equation} \label{eq:ilim_maglim}
    17.5 < i < 18 + 4 \times z_{\rm MEAN}
\end{equation}

An additional selection was applied in Y6, which includes a new Near Infra-Red (NIR) star-galaxy separation approach, using data from unWISE \citep{wise1}, and a non-parametric Self Organising Map (SOM)-based approach to identify regions in high-dimensional color-space with large amounts of systematic contamination from interlopers. We defer the reader to \citetalias{y6-maglim} for a more detailed description of such selections. 
We apply the joint mask \citepalias{y6-mask} resulting from the shear and LSS masks to both catalogs merged at \textsc{HealPIX} resolution  \textsc{Nside}=16384, which results in 4031 deg$^2$. We use the \textsc{Healsparse}\footnote{\texttt{https://healsparse.readthedocs.io/en/stable/}} software to create and apply such mask. 

The tomographic bin edges for the \maglimpp lens sample are defined using DNF output $z_{\rm MEAN}$ with the bin edges [0.2, 0.4, 0.55, 0.7, 0.85, 0.95, 1.05], matching those adopted in DES~Y3. The number of objects and number density per tomographic bin is reported in Table \ref{tab:data}, and plotted in Figure \ref{fig:nz}.

While six tomographic bins are defined for the \textsc{MagLim++} lens sample, we exclude bin 2 from our fiducial analysis. This decision follows findings from the full $3\times2$pt analysis, where lens bin 2 was found to contribute disproportionately to internal tensions, and the posterior of the leading redshift nuisance parameter of bin 2 showed significant deviations from the priors. A set of investigations failed to identify a conclusive cause, and the bin was conservatively removed prior to unblinding. For further discussion, see \citetalias{y6-3x2pt}.

\paragraph*{Photometric Redshifts}
The redshift distributions are calibrated using a combination of the \textsc{SOMPZ} and \textsc{WZ} methodologies developed for DES~Y6. The Self-Organising Maps Photo-Z (SOMPZ) method (\citepalias{y6-lenspz}) employs SOMs to transfer redshift information from deep multi-band fields with extensive spectroscopic coverage to the wide-field data, mapping color-magnitude space to redshift. This provides an initial estimate of the tomographic redshift distributions, $n(z)$. The Clustering Redshifts (WZ) method (\citepalias{y6-wz}) independently constrains the redshift calibration by measuring the angular cross-correlations between the lens samples and spectroscopic reference galaxies. The two estimates are then combined using an importance-sampling framework (\citepalias{y6-nzmodes}), which reweights the \textsc{SOMPZ} realizations according to their consistency with the \textsc{WZ} measurements. The resulting ensemble of $n(z)$ realizations is further compressed into a set of principal components, or modes, that capture the dominant sources of redshift uncertainty, ranked by their $\chi^2$ contribution to the 2pt data-vector. For \maglimpp, only 3 modes per tomographic bin are necessary to describe the data-vector variance. The amplitudes of these modes are treated as nuisance parameters and marginalized over in the cosmological inference. The reconstructed redshift distributions are showed in Figure \ref{fig:nz}.

\subsection{Source sample: \textsc{Metadetection}}

The full DES Y6 shape catalog described in \citetalias{y6-metadetect} covers 4422 deg$^2$. Applying the joint LSS-shear mask discussed above reduces the usable footprint to 4031 deg$^2$. The catalogue is constructed from cell-based image coaddition and shear measurements with Metadetection. 

Metadetection is an extension of the metacalibration framework that applies artificial shears directly to the detection process rather than only to measured objects. In this approach, detection, deblending, and measurement are recalculated on sheared versions of the image, thereby capturing selection effects that arise during object detection. This enables unbiased shear calibration even in crowded fields where metacalibration can fail.
The DES Y6 Metadetection weak lensing shape catalogue used in this work consists of 139,662,173 galaxies, with an effective number density of $n_{\rm eff}$ = 8.29 galaxies per arcmin$^2$ and shape noise of $\sigma_{\rm e}$ = 0.289. Additional per-tomographic-bin sample characteristics are reported in Table \ref{tab:data}.

\paragraph*{Photometric Redshifts}
The same methodology used for the lens galaxies is adopted also for the source galaxies. The redshift distributions are obtained from the \textsc{SOMPZ} \citepalias{y6-sourcepz} and \textsc{WZ} \citepalias{y6-wz} combination via importance sampling, with uncertainties encoded through the same set of principal modes used in the cosmological analysis. The only difference is that a single Self-Organizing Map is used to represent the wide-field color space, and the tomographic bins are defined directly as regions of this SOM rather than by external selection. This ensures a fully data-driven redshift calibration across all source bins. For the same reason, only 7 modes in total are needed, without the need to separate the compression by tomographic bin. The reconstructed redshift distributions, along with the corresponding lensing kernel, can be seen in Figure \ref{fig:nz}.

\section{Modelling the tangential shear}
\label{sec:modelling}

In this section we provide a brief overview of the main components of our modelling framework. A comprehensive description, including validation and alternative choices, is presented in \citetalias{y6-methods}.

The galaxy-galaxy lensing (GGL) signal quantifies the statistical correlation between the projected number density of lens galaxies and the mass distribution, as traced by the tangential shear of source galaxies around them.
For a lens tomographic bin $i$ and a source tomographic bin $j$, the observable is the tangential shear correlation function,
\begin{equation} \label{eq:gammat_definition}
\gamma_t^{ij}(\theta)
= \sum_{\ell=2}^{\infty} \frac{2\ell+1}{4\pi}
\frac{P_\ell^{2}(\cos\theta)}{\ell(\ell+1)}
C^{ij}_{\delta_{\rm g} E}(\ell),
\end{equation}
where $P_\ell^{2}$ are the associated Legendre polynomials and $C^{ij}_{\delta_{\rm g} E}(\ell)$ is the angular cross-power spectrum between the galaxy overdensity field in bin $i$, $\delta_{\rm g}$, and the E-mode shear field in bin $j$. Our intrinsic-alignment model can generate B-modes, but these contribute only to shear auto-correlations and thus not to the $1\times2$ galaxy-galaxy lensing signal, therefore we assume vanishing B-modes, so that the shear E-mode is equivalent to the convergence, $C^{ij}_{\delta_{\rm g} E} \equiv C^{ij}_{\delta_{\rm g} \kappa}$.

\vspace{0.3em}
Under the Limber approximation, the cross-spectrum is given by
\begin{equation}
C^{ij}_{\delta_{\rm g} \kappa}(\ell)
= \int d\chi
\frac{q^i_{\delta_{\rm g}}(\chi)q^j_\kappa(\chi)}{\chi^2}
P_{\rm gm}\left(k=\frac{\ell+1/2}{\chi}, z(\chi)\right),
\end{equation}
where $P_{\rm gm}(k,z)$ is the nonlinear matter power spectrum at redshift $z$ and comoving wavenumber $k=(\ell+1/2)/\chi$. The radial kernels are
\begin{equation}
q^i_{\delta_{\rm g}}(\chi)
= n^i_{\rm l}(z(\chi))\frac{dz}{d\chi}
\end{equation}
\begin{equation}
q^j_\kappa(\chi)
= \frac{3H_0^2\Omega_{\rm m}}{2c^2}
\frac{\chi}{a(\chi)}
\int_\chi^{\chi_{\rm H}} d\chi'
n^j_{\rm s}(z(\chi'))\frac{dz}{d\chi'}
\frac{\chi'-\chi}{\chi'} .
\end{equation}
Here $n^i_{\rm l}$ and $n^j_{\rm s}$ are the normalized redshift distributions of lens and source galaxies, respectively. In the linear bias case, $P_{\rm gm}(k,z)=b^i P_{\rm m}(k,z)$ with scale-independent galaxy bias $b^i$, while for nonlinear or scale-dependent bias we use the full perturbative expression for $P_{\rm gm}(k,z)$ described in \citetalias{y6-methods}.

We compute the linear matter power spectrum with \textsc{CAMB} \cite{Lewis_2000, Lewis_2002} and adopt \textsc{HMCode2020} \cite{Mead_2021} for the nonlinear regime. 
On the scales to which our data are most sensitive, baryonic feedback further suppresses power relative to dark-matter-only predictions, although the magnitude of this effect remains uncertain. Because of this, the first approach is to exclude scales where this effect would be significant. Baryonic feedback is incorporated through the sub-grid heating parameter $\log T_{\rm AGN}$, which controls the energy injected by AGN into the surrounding gas. The fiducial value $\log T_{\rm AGN}=7.7$ corresponds to an intermediate feedback strength consistent with the \textsc{BAHAMAS} simulations. This prescription is consistent with percent-level agreement with simulation-calibrated emulators on the scales probed by DES~Y6. Alternative models and validation are discussed in the companion modelling paper.

\paragraph*{Galaxy bias}

In the fiducial model we assume a linear, scale-independent galaxy bias, so that the galaxy--matter cross-power spectrum is related to the matter power spectrum by
\begin{equation}
    P_{\rm gm}(k,z) = b_1\,P_{\rm mm}(k,z),
\end{equation}
where $b_1$ is the linear bias parameter for the lens galaxy sample.
This approximation is expected to hold on the large, quasi-linear scales used in our fiducial analysis, and is sufficient for unbiased cosmological inference given our scale cuts.

We also explore a model that includes nonlinear and non-local bias terms, see Eqs.~from 15-19 from \citetalias{y6-methods}. In this perturbative framework, the galaxy overdensity receives contributions not only from the linear matter overdensity but also from quadratic, tidal, and higher-derivative terms, capturing the fact that galaxy formation does not trace the matter field in a purely local or linear fashion.
These nonlinear terms modify the galaxy-matter correlation on small and intermediate scales and allow us to test the sensitivity of the galaxy-galaxy lensing signal to departures from linear bias.

A reduced form of this perturbative model, commonly used in quasi-linear analyses, retains only the leading nonlinear correction, typically parameterized by a quadratic bias coefficient. In this case each lens bin is described by two free parameters, $(b_1^i, b_2^i)$, corresponding to the linear and quadratic bias terms \citep{Pandey2020}.
This simplified parametrization captures the dominant nonlinear deviations while limiting model complexity and the number of nuisance parameters.

% Following
% \citet{McDonald2009,Chan2012,Baldauf2012}, the galaxy--matter
% cross-spectrum can then be written as
% \begin{equation}
%     \begin{aligned}
%         P_{\rm gm}(k,z) &=  b_1\,P_{\rm mm}(k,z)
%         +  \frac{1}{2}b_2\,P_{b_1 b_2}(k,z)
%         +  \frac{1}{2}b_{s^2}\,P_{b_1 s^2}(k,z) \\
%         &\quad
%         + \frac{1}{2}b_{3\mathrm{nl}}\,P_{b_1 b_{3\mathrm{nl}}}(k,z)
%         + b_k\,k^2 P_{\rm mm}(k,z),
%     \end{aligned}
% \end{equation}
% where $b_2$, $b_{s^2}$, and $b_{3\mathrm{nl}}$ quantify second- and
% third-order local and nonlocal bias contributions, and $b_k$ represents
% a higher-derivative (scale-dependent) term. The cross-spectra
% $P_{b_1 b_2}$, $P_{b_1 s^2}$, and $P_{b_1 b_{3\mathrm{nl}}}$ are
% one-loop integrals between the linear and higher-order bias operators.
% This extended model allows us to test the impact of nonlinear biasing on
% the galaxy--galaxy lensing signal and to verify that our fiducial linear
% bias treatment is adequate on the scales considered.

% On large scales the lens galaxies are related to the underlying matter field through a linear bias parameter $b_1^i$ in each redshift bin. On smaller, quasi-linear scales, we allow for additional nonlinear contributions. Our fiducial analysis includes two free bias parameters per lens bin, $(b_1^i, b_2^i)$, corresponding to the linear and quadratic terms in a perturbative expansion \cite{Pandey_2020}. This provides sufficient flexibility to capture deviations from linear bias while keeping the number of nuisance parameters tractable.

\paragraph*{Scale cuts}\label{sec:scalecuts}
Although the galaxy--galaxy lensing signal is measured over the full range \(2.5 < \theta \lesssim 1000\) arcminutes, our theoretical modeling is not reliable across all scales. We therefore apply scale cuts to exclude angular ranges where unmodeled physical effects could bias the inferred cosmological parameters. 

At small angular separations, the signal can be affected by complex astrophysical and observational processes--such as baryonic feedback and the limitations of the point-mass term, which approximates all unresolved small-scale mass as a single monopole and therefore cannot fully capture the true scale-dependent departures from the halo-model prediction. To determine the minimum angular scales to retain, we follow the procedure adopted in the DES~Y3 cosmological analysis \citep{y3-3x2ptkp}. We generate simulated data vectors that include the nonlinear power spectrum and are then contaminated with small-scale effects (e.g., baryonic feedback). We use these contaminated vectors to perform the full cosmological inference (see \citetalias{y6-methods}). The resulting posteriors are compared to those obtained from uncontaminated (fiducial) data vectors, and the minimum angular scale for each lens redshift bin is chosen such that the bias in the recovered cosmological parameters \((\Omega_m, S_8)\) remains smaller than \(0.3\sigma\). This procedure leads to a comoving minimum scale of \(R_{\min}=6\,h^{-1}\mathrm{Mpc}\), which translates to angular cuts of \(24.3', 17.6', 13.0', 10.9', 9.7'\), and \(8.7'\) for the six lens tomographic bins, respectively. 
For analyses adopting a nonlinear galaxy bias prescription, we define an alternative set of cuts based on a comoving threshold of \(R_{\min}=4\,h^{-1}\mathrm{Mpc}\), which results in angular cuts of \(16.2', 11.8', 8.7', 7.2', 6.4'\), and \(5.9'\). This ensures that all scales included in the analysis lie safely within the regime where higher-order bias terms and nonlinear effects remain well modeled.

At large separations, the measurements become increasingly sensitive to survey geometry, masking, and residual systematics in the random catalogs, which are hard to characterize. For Y6, we conservatively limit the analysis to $\theta_{\max} = 250'$, although measurements up to $\sim 1000'$ are available and may be incorporated in future analyses once the modelling of large-scale systematics and covariances is further refined. This choice is motivated by tests in \citetalias{y6-methods}, which showed that extending the data vector to the full measured range ($\sim 1000'$) yields only a marginal improvement in cosmological constraints in a simulated cosmic-shear analysis, and no compelling justification--such as demonstrably well-controlled large-scale systematics across all probes--was identified to warrant the substantial additional modelling and validation effort required to include those scales.

\begin{figure*}
    \centering
    \includegraphics[width=\textwidth]{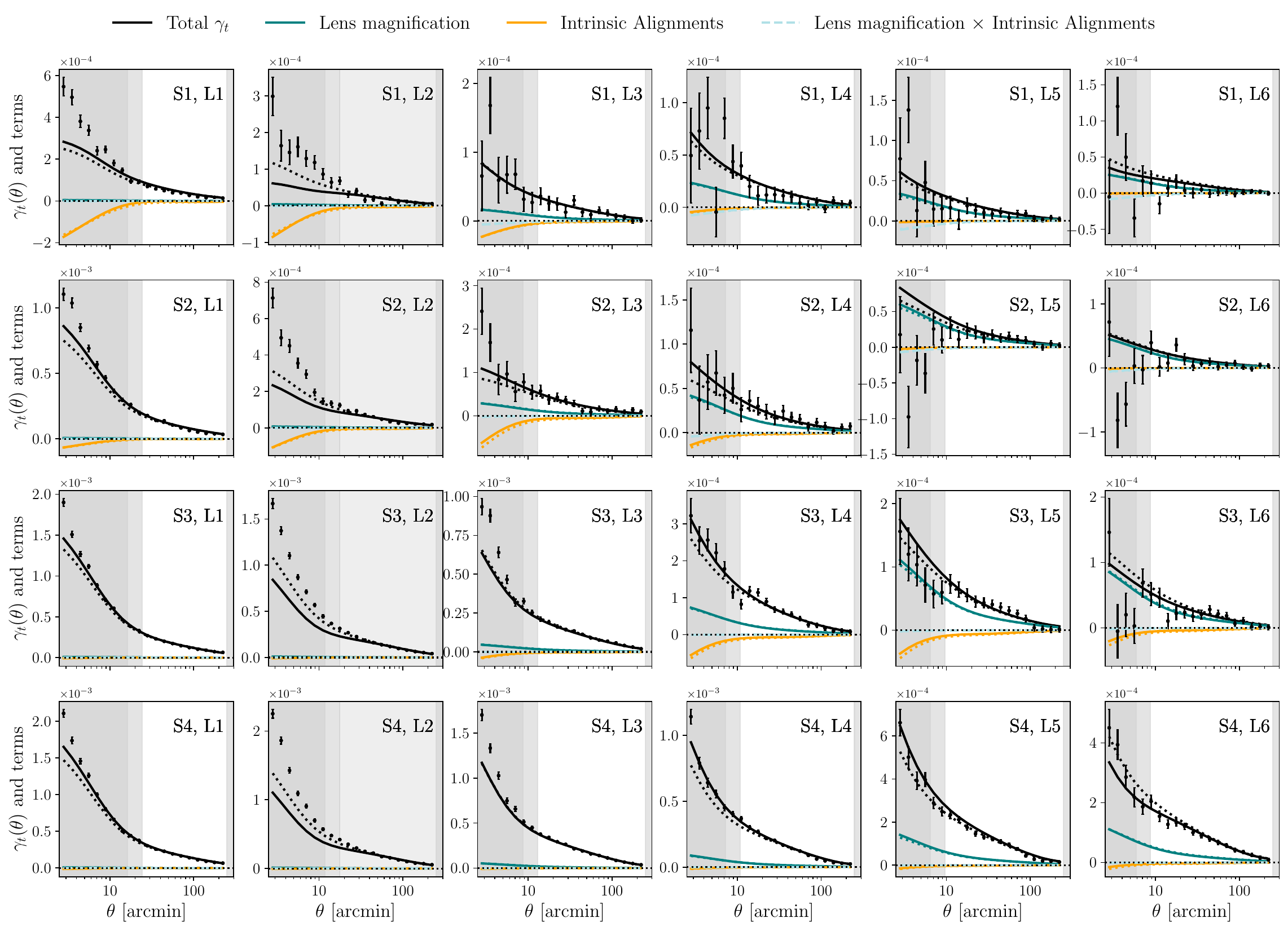}
    \caption{Contributions to the tangential shear signal from the different components of the fiducial model, evaluated at the best-fit parameters of the DES Y6 3$\times$2pt analysis, shown for each lens-source bin pair. Solid curves correspond to the prediction adopting linear galaxy bias, while dashed curves show the prediction with nonlinear bias. The black curve denotes the total model prediction, and the individual contributions from intrinsic alignments, lens magnification, and their cross term are shown separately. The black points (with error bars) represent the measured tangential shear. Gray shaded regions mark the angular scales removed from the analysis: darker gray for the linear bias scale cuts and lighter gray for the nonlinear bias cuts. Lens bin 2 is fully shaded because it is excluded from the fiducial data vector. Note that this figure does not illustrate the point-mass marginalization, which is applied during likelihood evaluation through the inverse covariance matrix and therefore cannot be represented in terms of the covariance elements shown here. }
    \label{fig:contrib}
\end{figure*}

\paragraph*{Intrinsic alignments and lens magnification}

Our fiducial model further accounts for additional contributions to the
observed shear signal arising from intrinsic alignments (IA) of galaxies
and from lens magnification. In the presence of these effects, the total
angular cross-power spectrum between the lens galaxy density and the
observed E-mode shear field becomes
\begin{equation}
C^{ij}_{\delta_{\rm g} E}(\ell)
= C^{ij}_{\delta_{\rm g} \kappa}(\ell)
+ C^{ij}_{\delta_{\rm g} I_E}(\ell)
+ C^{ij}_{\delta_{{\rm \mu}} \kappa}(\ell)
+ C^{ij}_{\delta_{{\rm \mu}} I_E}(\ell),
\end{equation}
where $\delta_\mu$ denotes the magnification contribution to the observed lens galaxy density, and $I_E$ represents the E-mode component of the intrinsic alignment field. 

The first term, $C^{ij}_{\delta_{\rm g} \kappa}$, represents the standard galaxy--galaxy lensing contribution, while $C^{ij}_{\delta_{\rm g} I_E}$ accounts for the cross-correlation between the lens density and intrinsic galaxy shapes (the ``GI'' term). The remaining two terms arise from lens magnification: $C^{ij}_{\delta_\mu \kappa}$ describes correlations between magnified lens counts and the gravitational shear, and $C^{ij}_{\delta_\mu I_E}$ represents the mixed magnification--IA contribution. In the absence of magnification and intrinsic alignments, only $C^{ij}_{\delta_{\rm g} \kappa}$ remains.

Intrinsic alignments (IA) of galaxy shapes with the tidal field can mimic lensing-induced correlations. We model these effects using both the nonlinear linear alignment (NLA) model \citep{Bridle_2007, hirata2004} and the more general tidal alignment and tidal torquing (TATT) model \citep{Blazek_2019}. For our fiducial $2\times2$pt and $3\times2$pt analyses we adopt the TATT-4 parameterization, in which the density-weighting parameter is fixed to $b_{ta}=1$. The detailed implementation and validation of these models is described in \citetalias{y6-methods}.
In Appendix~\ref{app:ggl_ia} we provide a detailed discussion of how the tangential shear measurement constrains the intrinsic-alignment parameters of the TATT model.

Gravitational lensing modifies the observed density of lens galaxies through magnification. The effect is parameterized by an effective magnification coefficient
\begin{equation}\label{eq:magnification}
    C_{\rm sample} = 2(\alpha - 1),
\end{equation}
where $\alpha$ captures how the number of observed, selected galaxies responds to lensing, including selection effects. We adopt Gaussian priors on $\alpha$, derived from Balrog image simulations and presented in \citetalias{y6-magnification, y6-balrog}.

The different contributions from IA and magnification, shown in Figure \ref{fig:contrib}, are evaluated at their best-fit values. Intrinsic alignments yield a negative contribution for redshift-overlapping bins, consistent with the expected anti-correlation between density and tangential shape. Magnification, on the other hand, becomes increasingly important at higher lens redshift and can exceed the IA term in several bins. The mixed magnification-IA term remains negligible in all cases. Although IA and magnification can each reach non-trivial amplitudes depending on the bin pair, their combined impact is still subdominant relative to the total shear signal; including them therefore improves modeling accuracy without qualitatively changing the inferred signal amplitude.

% The corresponding radial kernels are then defined as
% \begin{align}
%     q^i_{\delta_g}(\chi)
%         &= b^i\,n^i_{\rm l}(z(\chi))\,\frac{dz}{d\chi}, \\
%     q^i_{\delta_\mu}(\chi)
%         &= 2\!\left(\alpha^i - 1\right)
%            \frac{3H_0^2\Omega_{\rm m}}{2c^2}
%            \frac{\chi}{a(\chi)}
%            \int_\chi^{\chi_{\rm H}} d\chi'\,
%            n^i_{\rm l}(z(\chi'))\,\frac{dz}{d\chi'}\,
%            \frac{\chi' - \chi}{\chi'}, \\
%     q^j_\kappa(\chi)
%         &= \frac{3H_0^2\Omega_{\rm m}}{2c^2}
%            \frac{\chi}{a(\chi)}
%            \int_\chi^{\chi_{\rm H}} d\chi'\,
%            n^j_{\rm s}(z(\chi'))\,\frac{dz}{d\chi'}\,
%            \frac{\chi' - \chi}{\chi'}, \\
%     q^j_{I_E}(\chi)
%         &= -\,A_{\rm IA}\,C_1\,\rho_{\rm crit}
%            \frac{\Omega_{\rm m}}{D(z(\chi))}\,
%            n^j_{\rm s}(z(\chi))\,\frac{dz}{d\chi}.
% \end{align}
% Here $\rho_{\rm crit}$ is the critical density at $z=0$,
% $D(z)$ is the linear growth factor normalized to unity today, and
% $A_{\rm IA}$ and $C_1$ set the overall amplitude and normalization of
% the IA model, respectively. The negative sign in $q^j_{I_E}$ reflects
% the tendency of galaxies to align radially toward overdensities.

\paragraph*{Point-mass marginalization}\label{par:pm}

Because the tangential shear at a given angular separation $\theta$ receives contributions from the mass distribution on all smaller scales, inaccuracies in modeling the halo-matter correlation on small scales can bias large-scale measurements. To mitigate this non-locality, we adopt point-mass marginalization \citep{MacCrann2020, Prat2023}, which models unresolved small-scale mass as an additive scaling term $A^{ij}/\theta^2$ 
\begin{equation}
    \gamma_t^{ij}(\theta) \;\rightarrow\; \gamma_{t,\mathrm{model}}^{ij}(\theta) + \frac{A^{i}}{\theta^2},
\end{equation}
and analytically marginalizes over its amplitude by modifying the inverse covariance matrix. We adopt the thin-lens approximation, treating the lens bin as a single effective plane by assuming that the point-mass amplitude varies only slowly across the finite redshift width of the bin. As a result, the enclosed unresolved mass can be represented by one parameter per lens bin. In this limit, the relative contribution of this point-mass term to different source bins is fully determined by geometric lensing-efficiency factors and can therefore be computed analytically. Although this neglects small variations of the enclosed mass across the bin, DES Y3 tests \citep{Prat2023} showed that these effects are negligible for bins of similar width, validating the use of a single parameter per lens bin without introducing biases.

% The tangential shear signal is non-local: small-scale uncertainties can propagate to larger radii. We account for this with point-mass (PM) marginalization \cite{MacCrann2020}, which introduces a term with $1/\theta^2$ scaling,
% \begin{equation}
%     \gamma_t^{ij}(\theta) \;\rightarrow\; \gamma_{t,\mathrm{model}}^{ij}(\theta) + \frac{A^{i}}{\theta^2}.
% \end{equation}
% We marginalize analytically over one PM amplitude per lens bin, removing sensitivity to scales below our chosen cut while retaining unbiased constraints \cite{Prat2023}.

\paragraph*{Redshift calibration}

We calibrate the redshift distributions of both lens and source samples using mode projection \citepalias{y6-nzmodes}, in which each distribution is modeled as a mean $n(z)$ plus a linear combination of basis modes. The amplitudes of these modes are treated as nuisance parameters and marginalized over in the inference. This approach captures both photometric and clustering-redshift uncertainties \citepalias{y6-lenspz, y6-sourcepz, y6-wz}. 

We reconstruct the redshift distributions using a mode projection approach, marginalizing over 7 modes \( U_m(z) \) shared across the 4 source redshift bins:
\begin{equation}
  % n^{\rm i}(z) = \bar n^i(z) + \sum_{\rm m} u^{\rm m}\, U^{\rm i,m}(z),
 n^i_{\rm s}(z) =  \bar n_{\rm s}^i(z) + \sum_{\rm m=1}^{7}  u_{\rm s}^{\rm m}\, U_{\rm s}^{i,\rm m}(z),
 % \bar{n}^i_s(z) + \sum_{m=1}^{7} u^s_{m}\, U_{m}(z),
 \label{eq:modes_s}
\end{equation}
and over 3 modes \textit{per} lens redshift bin:
\begin{equation}
 n^i_{\rm l}(z) =  \bar n_{\rm l}^i(z) + \sum_{\rm m=1}^{3}  u_{\rm l}^{\rm m}\, U_{\rm l}^{\rm m}(z),
 % n^i_l(z) = \bar{n}^i_l(z) + \sum_{m=1}^{3} u^l_{im}\, U_{im}(z),
  \label{eq:modes_l}
\end{equation}
where \( \bar{n}_{\rm l/s}(z) \) denotes the mean redshift distribution for lenses or sources, respectively. The mode amplitudes \( u \) are scalar coefficients drawn from standard normal distributions and marginalized over during inference.
The reason we consider a separate set of modes for each lens tomographic bin is that the lens-bin definition is fixed a priori (unlike for the source galaxies, whose binning is finalized during the redshift calibration). For this reason, we treat the lens bins as independent throughout the entire redshift-calibration procedure.

\paragraph*{Shear calibration}

Finally, we account for uncertainty in shear calibration by marginalizing over a multiplicative bias parameter $m^i$ in each source bin, defined by
\begin{equation}
    e_j^i = (1+m^i)\,\gamma_j^i,
    \label{eq:addm}
\end{equation}
with Gaussian priors informed by Balrog simulations.

\subsection{Covariance}\label{sec:theorycov}

The covariance of the $3\times2$pt data vector is computed analytically using \textsc{CosmoCov} \citep{Krause2017, Fang2020}, which includes both Gaussian and non-Gaussian terms and was extensively validated for DES Y3 \citep{y3-covariances}. As in Y3, we incorporate analytic marginalization over uncertainties in the observational systematics weights that enter the clustering estimator $w(\theta)$; this marginalization produces an additional systematic covariance that is added directly to the analytic prediction. For galaxy-galaxy lensing specifically, we also account for the point-mass (PM) contribution, although in this case the marginalization is implemented at the level of the inverse covariance matrix during likelihood evaluation and therefore cannot be expressed as a standard additive covariance term.

A small number of updates in Y6 modify the covariance relative to Y3 in controlled and well-understood ways. First, the adoption of non-zero central values for the multiplicative shear biases requires adjusting the effective source galaxy number densities that enter the shear and galaxy-galaxy lensing covariance components. Second, while the Y6 signal modeling uses \textsc{HMCode2020}, the covariance calculation continues to employ \textsc{Halofit} \citep{Takahashi2012} for the nonlinear matter power spectrum. This mismatch results in differences of at most $\sim 2.5\%$ in individual covariance elements and yields negligible changes in the resulting parameter uncertainties. Finally, survey-geometry corrections to the shape- and shot-noise terms are recomputed directly in configuration space using the Y6 random catalogs, a refinement necessitated by the increased small-scale masking and spatially varying completeness of Y6 relative to Y3. These corrections primarily affect the noise component of the galaxy-galaxy lensing covariance.

\section{Measurement: tangential shear estimator}
\label{sec:measurement}

\begin{figure*}
    \centering
    \includegraphics[width=\textwidth]{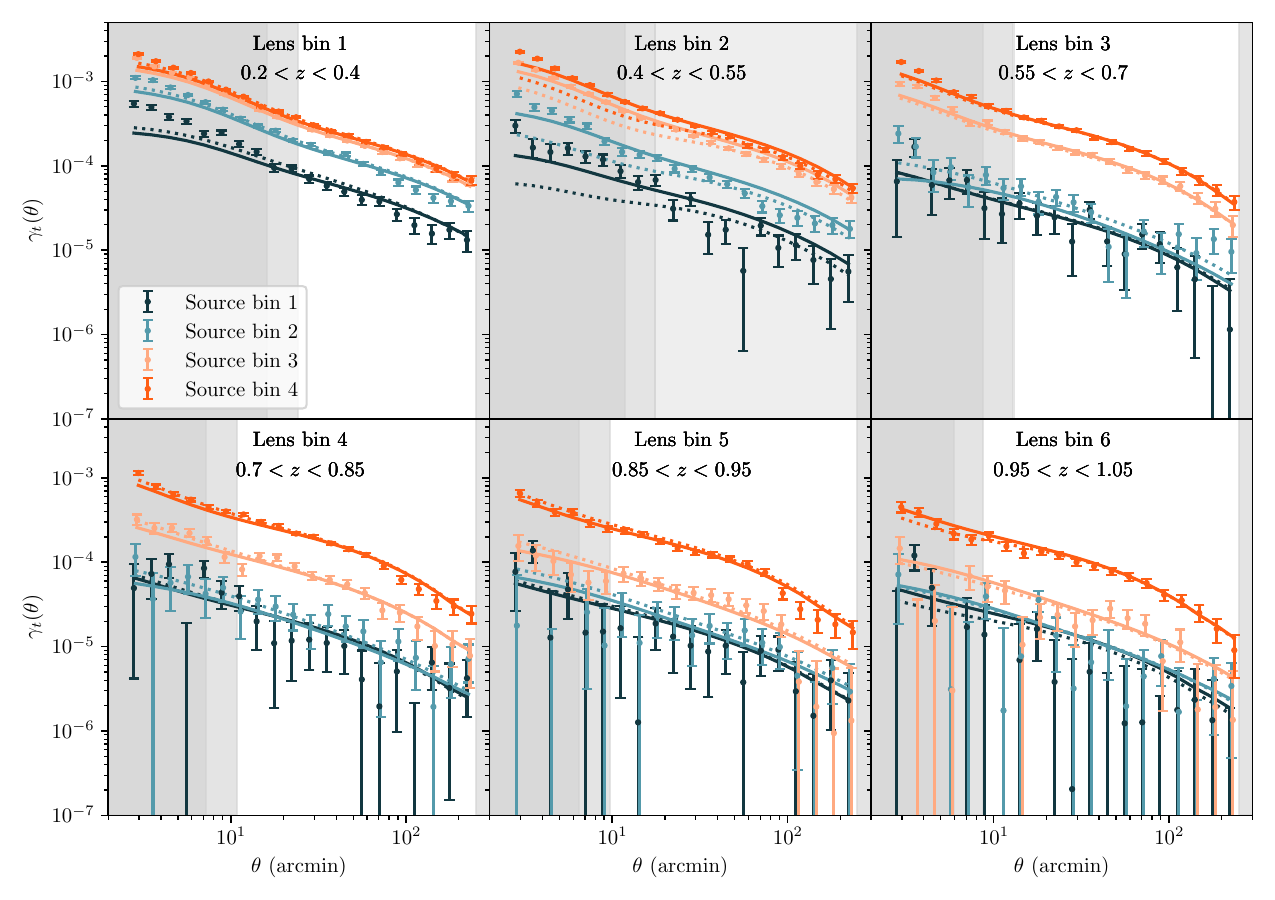}
    \caption{Tangential shear measurement for the 6 tomographic bins of MagLim++ and the 4 tomographic bins of \mdet. We overlay the best fit model for the linear bias case (solid lines) and for the nonlinear bias case (dashed lines). Gray shaded regions mark the angular scales removed from the analysis: darker gray for the linear bias scale cuts and lighter gray for the nonlinear bias cuts. Lens bin 2 is fully shaded because it is excluded from the fiducial data vector. Note that this figure does not illustrate the point-mass marginalization, which is applied during likelihood evaluation through the inverse covariance matrix and therefore cannot be represented in terms of the covariance elements shown here. }
    \label{fig:gammat}
\end{figure*}

We now describe how the galaxy-galaxy lensing signal introduced in the previous section is measured from the data. We first outline the measurement pipeline, and then present the baseline estimator used to relate source ellipticities to the tangential shear. We also discuss the corrections required by our specific measurement procedure, and conclude with the final form of the corrected estimator used in the analysis.

\subsection{Measurement Pipeline}

The measurement is carried out using \textsc{TreeCorr} v5.1 \citep{treecorr}, a publicly available two-point correlation function code that supports large-scale parallel processing and accounts for survey geometry, masking, and angular completeness. Both the lens and source catalogs are divided into spatial patches as required by \textsc{TreeCorr}, and the tangential shear signal is computed for all lens-source tomographic bin pairs.

We use 26 logarithmically spaced angular bins spanning the range $2.5^\prime < \theta \lesssim 1000^\prime$ , which encompasses both the quasi-linear and nonlinear regimes of structure formation. Only the 20 bins between $2.5^\prime < \theta < 250^\prime$ are  used for the main analysis.  The application of scale cuts to remove scales dominated by poorly modeled systematics is described in Section~\ref{sec:scalecuts}.

TreeCorr accelerates correlation function calculations by grouping objects into cells and using cell-center approximations for the pair separation and, for shear quantities, the projection angle. The parameters \texttt{bin\_slop} and \texttt{angle\_slop} set the maximum tolerated inaccuracies in these approximations. The \texttt{bin\_slop} parameter controls the allowed fractional error in the separation relative to the bin width, and \texttt{bin\_slop}=0 forces TreeCorr to place every pair into its exact separation bin without any tolerance. The \texttt{angle\_slop} parameter limits the maximum allowed difference between the approximate and true orientation of a pair used for shear projections; \texttt{angle\_slop}=0 enforces exact pair orientations. While these choices increase the computational cost, they ensure that the separation binning and shear projections are evaluated with maximum accuracy.

\paragraph*{Jackknife Covariance}\label{sec:jackknifecov}
To interpret the measurements and assess their statistical significance, it is necessary to estimate the associated uncertainties in the form of a covariance matrix. In this work, we rely on both theoretical predictions (see Section~\ref{sec:theorycov}) and data-driven estimates based on jackknife resampling. The jackknife method estimates statistical uncertainties by recalculating the measurement multiple times, each time excluding one spatial region of the survey, and computing the covariance from the variance across these realizations. In our implementation, the jackknife regions are generated directly by \textsc{TreeCorr}. We provide \textsc{TreeCorr} with the random catalog and specify the desired number of jackknife patches ($N_{\mathrm{JK}} = 250$), allowing the software to partition the survey footprint based on the spatial distribution of the random points. The resulting patch centres are saved and reused when constructing the lens and source catalogs, ensuring a consistent definition of jackknife regions across all fields. The number of patches is chosen such that the maximum angular scale used in the analysis ($\theta_{\max} = 250'$) fits well within a single jackknife region, and the jackknife covariance is therefore computed only up to this scale. The resulting pipeline ensures consistent propagation of survey geometry, masking, and estimator-level effects throughout the analysis.

\subsection{Baseline Estimator}

The following describes the construction of the lens-source pairs, the definition of the tangential ellipticity component, and the estimators used to compute the mean shear profile as a function of angular separation.

For a given lens-source pair ($LS$), the tangential component of ellipticity as a function of the two components of ellipticity reads as: 
\begin{equation}
    e_{t, \rm LS} = -e_1 \cos(2\phi) + e_2 \sin(2\phi),
    \label{e_tangential}
\end{equation}
where $\phi$ is the angle between the horizontal cardinal axis and the line connecting the galaxy pair.

We can now use the tangential ellipticity as a proxy for the tangential shear:
\begin{equation}
    \gamma_t = \frac{\sum_{\rm LS} w_{\rm LS} e_{t,\rm LS}(\theta) }{\sum_{\rm LS} w_{\rm LS}(\theta)}\label{eq:baseline_estimator}
\end{equation}

where $w_{\rm LS} = w_{\rm l} w_{\rm s}$ is the weight factor for a given lens-source pair. $w_{\rm l}$ is the lens galaxy clustering weight, while $w_{\rm s}$ is the source weight. Lens clustering weights are designed to correct for spatial variations in the observed lens density caused by both large-scale structure and observing conditions (e.g., depth, seeing, sky brightness). While such variations would not bias the galaxy-galaxy lensing signal if the lens and source samples were affected by uncorrelated systematics, in practice both are influenced by the same survey conditions. This shared dependence induces correlations between the observed densities of lenses and sources that can bias the measured tangential shear through spurious pair count modulations. Applying the lens weights therefore mitigates these correlated systematics and ensures that the measured lensing signal is not artificially amplified or suppressed. 
% In Figure \ref{fig:residuals_corrections} the impact of the clustering weights correction is shown in form of residual with respect to the fiducial measurement. 
The source's shear weights are computed to optimize the measurement of tangential shear by accounting for the varying contributions of individual sources. These weights are typically calculated as the inverse variance of the galaxy's ellipticity, which incorporates both the intrinsic shape noise (the natural dispersion in galaxy ellipticities) and measurement uncertainties. The weights are further scaled by the shear response, a factor that quantifies how the observed ellipticity changes in response to an applied shear, described in more detail in section \ref{sec:response}.

In practice, several observational, astrophysical, and statistical effects can bias this basic estimator of the tangential shear. The following subsections describe the methodology adopted to measure the signal and estimate its covariance, as well as the set of corrections applied to mitigate these effects, including boost factors, random-point subtraction, and shear-response, ensuring an unbiased estimate of the underlying galaxy-matter correlation.

% \begin{equation}
%     \gamma_t = \frac{1}{R}\frac{\sum_{\rm LS} w_{\rm LS} e_{t,\rm LS}(\theta) }{\sum_{\rm LS} w_{\rm LS}(\theta)}
% \end{equation}
% %

\subsection{Boost Factor}\label{sec:boosts}

\begin{figure*}
    \centering
    \includegraphics[width=\textwidth]{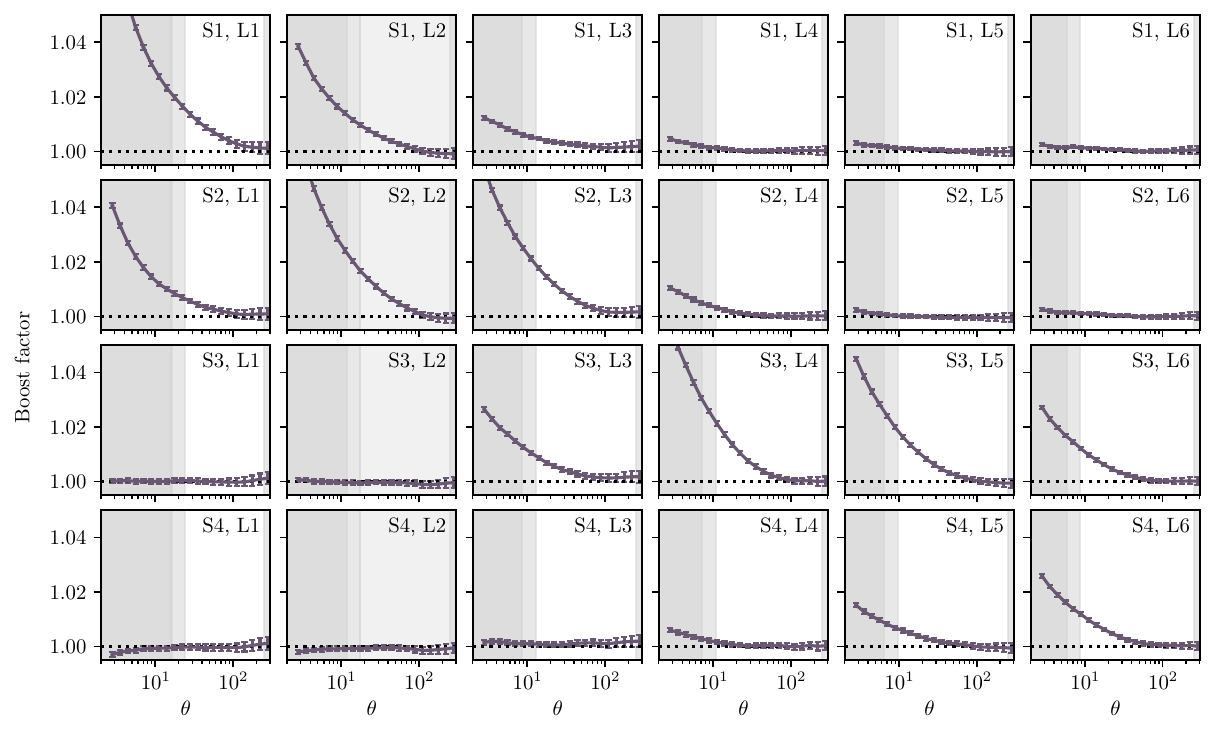}
    \caption{Boost factors with their uncertainty for the 6 tomographic bin of \maglimpp and the 4 bins of \mdet. Gray shaded regions mark the angular scales removed from the analysis: darker gray for the linear bias scale cuts and lighter gray for the nonlinear bias cuts. Lens bin 2 is fully shaded because it is excluded from the fiducial data vector. Note that this figure does not illustrate the point-mass marginalization, which is applied during likelihood evaluation through the inverse covariance matrix and therefore cannot be represented in terms of the covariance elements shown here. }
    \label{fig:boosts}
\end{figure*}

Our modeling assumes that the lens and source samples are statistically independent, such that source galaxies are not spatially correlated with the lenses. This assumption breaks down when the redshift distributions of the two samples overlap, resulting in a non-negligible fraction of lens-source pairs that are physically associated. These galaxies may reside in the same large-scale structure or halo as the lens, and are therefore not genuinely lensed. However, because they enter the pair counts used to estimate the shear, they dilute the measured signal relative to the model prediction, particularly on small angular scales, leading to a scale-dependent bias in the tangential shear measurement.

To correct for this effect, we apply a \textit{boost factor} correction, which quantifies the excess number of source galaxies physically associated with the lens sample, relative to what is expected for an random source distribution. The boost factor, \( B(\theta) \), is defined as:

\begin{equation}
B(\theta)\equiv 1+\omega_{\rm LS}(\theta)
= \frac{N_{\rm r}}{N_{\rm l}}\,
\frac{\displaystyle \sum\nolimits_{\rm (l,s)} w_{\rm l} w_{\rm s}}
{\displaystyle \sum\nolimits_{\rm (r,s)} w_{\rm s}}.
\label{eq:boosts}
\end{equation}
where $\omega_{\rm LS}(\theta)$ is the angular correlation function between lenses and sources, describing the excess probability of finding a source galaxy at an angular separation  from a lens galaxy compared to a random distribution, while $N_{\rm l}$ and $N_{\rm r}$ are the weighted number of lenses and random points, respectively.

We can then rewrite the tangential shear of Eq. \ref{eq:baseline_estimator} as:  
\begin{equation}
    \gamma_{t, \rm no bf (\theta)} =  \frac{\sum_{\rm l} w_{\rm l}}{\sum_{\rm r} w_{\rm r}} \frac{\sum_{\rm RS} w_{\rm RS}(\theta)}
    {\sum_{\rm LS}w_{\rm LS}(\theta)} \frac{\sum_{\rm r} w_{\rm r}}{\sum_{\rm l} w_{\rm l}} \frac{\sum_{\rm LS} w_{\rm LS}(\theta) e_{t, \rm LS}(\theta)}{\sum_{\rm RS} w_{\rm RS}(\theta)},
\end{equation}
where $w_{\rm RS}$ is the weight associated to each random-source pair, similarly as $w_{\rm LS}$. At this point it is easy to express the tangential shear estimator, corrected for the boost factors of Eq. \ref{eq:boosts}: 
\begin{equation}
    \gamma_{t, \rm bf (\theta)} =  \gamma_{t, \rm no bf}B(\theta) = \frac{\sum_{\rm r} w_{\rm r}}{\sum_{\rm l} w_{\rm l}} \frac{\sum_{\rm LS} w_{\rm LS}(\theta)e_{t, \rm LS}(\theta)}{\sum_{\rm RS}w_{\rm RS}(\theta)}.
\end{equation}

% The boost factor modifies the tangential shear estimator to correct for this clustering effect. Without the correction, the measured tangential shear would underestimate the true signal. 
By normalizing the lens-source pair counts using random-source pair counts, the boost factor effectively removes the clustering bias, ensuring that the measured tangential shear matches our modeling.

To estimate the covariance of the boost factor, we compute it directly from the same pair-count quantities that define the boost itself. For each jackknife resample, we evaluate the ratio between the weighted number of lens-source pairs and the corresponding random-source pairs, both normalized by the total lens and random weights in that patch. This produces a boost-factor estimate in each angular bin with one jackknife region removed. TreeCorr's multi-catalog jackknife capability then assembles these leave-one-out evaluations into a full covariance matrix for the boost factor. In this way, the boost-factor covariance is derived self-consistently from the underlying LS and RS pair counts and their patch-by-patch fluctuations, without introducing any additional approximations or surrogate quantities.

As shown in Figure \ref{fig:boosts}, the boost factor correction is most significant on small angular scales, where lens-source clustering is strongest. These corrections have a maximum impact of $\sim$ 7\% at the smallest measured angular scale, and of $\sim$ 2-3\% at the smallest scale included in the DES Y6 $3\times2pt$ cosmological analysis (equivalent to 6 Mpc/$h$ for the linear bias regime, and 4 Mpc/$h$ for the nonlinear bias regime, see Section \ref{sec:scalecuts}). The boost factor approaches 1 at large separations, confirming that these small-scale deviations are local and consistent with noise and survey geometry effects.

For a few lens-source bin combinations, the measured boost factor falls slightly below unity at small scales. Such deviations are expected: they typically reflect small mismatches in the completeness captured by the random catalog or shot-noise fluctuations in low-pair-count regimes. These modest deficits have only a minimal impact on the inferred tangential shear, and they occur on angular scales that are removed by our scale cuts.

Although the boost-factor correction is applied at all angular scales, the scale cuts remove exactly the regime in which the boost deviates most strongly from unity. As a result, the correction has a negligible influence on the retained tangential-shear measurements, which we examine in Section \ref{sec:impact}.

\begin{figure*}
    \centering
    \includegraphics[width=\textwidth]{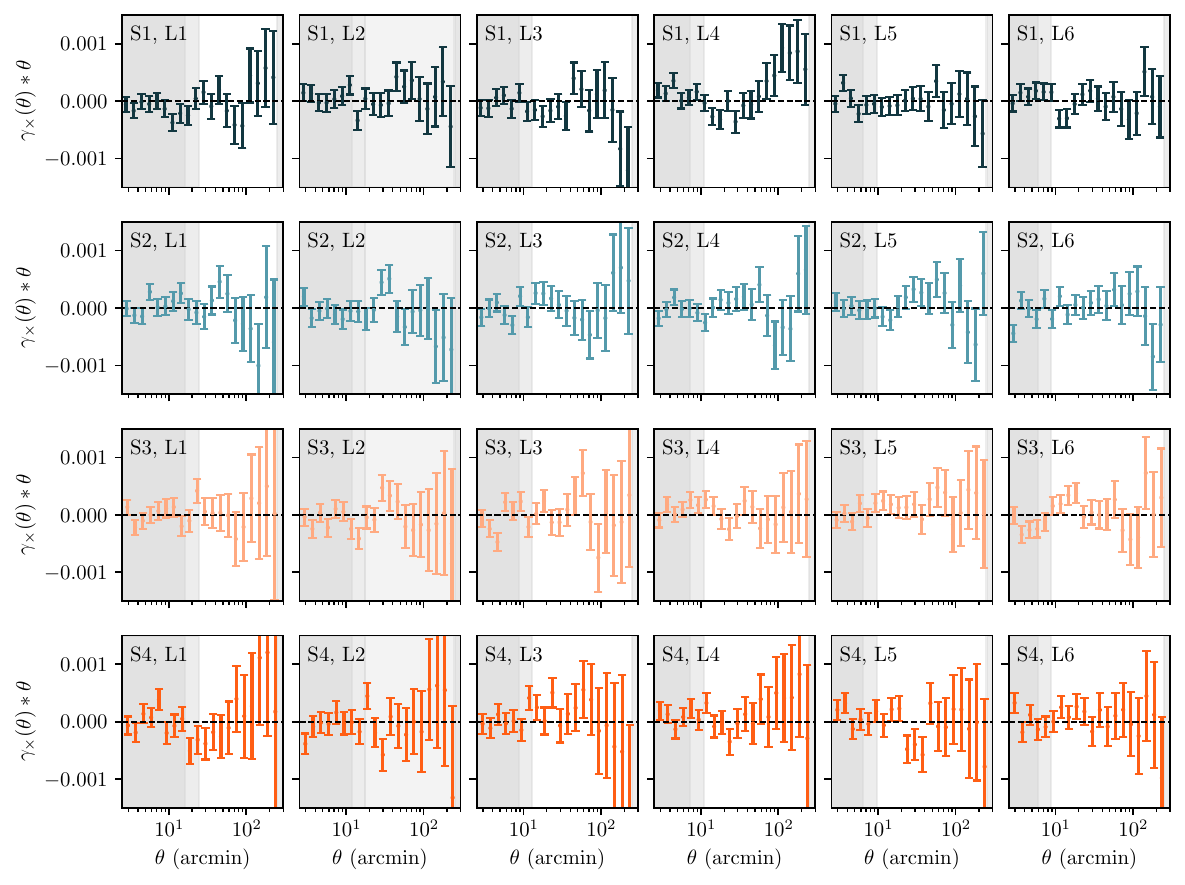}
    \caption{The cross-component of shear, $\gamma_\times(\theta)$, for each lens bin across six lens bins. The error bars are computed from the Jackknife covariance. Each row represents a different source bin combination, showing the behavior of $\gamma_\times(\theta)$ as a function of angular separation $\theta$ (arcmin). Gray shaded regions mark the angular scales removed from the analysis: darker gray for the linear bias scale cuts and lighter gray for the nonlinear bias cuts. Lens bin 2 is fully shaded because it is excluded from the fiducial data vector. }
    \label{fig:gammax}
\end{figure*}

\subsection{Random points subtraction}\label{sec:randoms}

A critical step in ensuring the robustness of this measurement is accounting for systematic effects related to survey geometry, masks, and selection biases. To address these issues, we perform the subtraction of signals measured around random points. Random points serve as a control sample that models the survey's response to systematic effects unrelated to the gravitational lensing signal (e.g., residual mean shear, selection anisotropies, and small residual mask- or depth-dependent shear systematics). Moreover, the subtraction of tangential shear measured around random points addresses systematic effects specific to the tangential coordinate system. While galaxy-shear cross-correlations are less sensitive to additive shear systematics compared to shear-shear correlations, this requires the sources to be isotropically distributed around the lens and a spatially constant additive shear. These assumptions often fail near survey edges or in heavily masked regions, where source distributions are asymmetric. Measuring the tangential shear around random points provides a robust way to remove these additive systematics, as random points better sample the survey edge and masked regions, ensuring systematic effects are effectively measured and subtracted.

In addition to mitigating survey systematics, the subtraction of the tangential shear around random points removes a distinct contribution to the covariance that arises when the estimator is expressed in terms of the overdensity field rather than the raw density field. As shown by \citep{Singh2017}, fluctuations in the survey mask and in the mean density couple nonlinearly to the overdensity field, generating an additional variance term that inflates the covariance. Subtracting the random-point signal eliminates this term, improving the precision of the measurement without changing its mean.

These covariance benefits are complementary to the removal of additive systematics discussed above. Together, they ensure that random subtraction not only protects the measurement from survey-dependent biases but also yields a more stable and reliable covariance estimate, which is essential for robust cosmological inference. 
We show in Appendix \ref{app:random_points} the tangential-shear signal measured around random points, as well as the impact on the jackknife covariance when this contribution is removed or retained.

% In addition to mitigating survey systematics, the subtraction of the tangential shear around random points also removes a covariance term introduced by the use of the overdensity field rather than the density field in the measurement. This contribution to the covariance matrix arises from the nonlinear nature of the overdensity field and its coupling to survey geometry. Subtracting the random points' signal significantly improves the accuracy of GGL measurements by removing large-scale systematics that could bias cosmological parameter estimation. 

% This step is particularly critical in modern photometric surveys, where complex masks and selection effects can introduce spurious signals. These arise from non-uniform angular distributions of lens galaxies caused by survey edges, gaps, and varying depth, as well as from spatial variations in shear calibration or intrinsic alignments that correlate with survey features. Without random subtraction, such effects can mimic or obscure the true gravitational lensing signal, leading to biased inferences. In addition, random subtraction improves the covariance estimation, yielding more reliable uncertainty estimates and enhancing the overall robustness of GGL analyses.

We can further modify the tangential shear estimator, by subtracting the systematic component from the true lensing-induced shear:
\begin{equation}
    \gamma_{t} = \frac{\sum_{\rm r} w_{\rm r}}{\sum_{\rm l} w_{\rm l}} \frac{\sum_{\rm LS} w_{\rm LS}(\theta)e_{t, \rm LS}(\theta)}{\sum_{\rm RS}w_{\rm RS}(\theta)} - \frac{\sum_{\rm RS} w_{\rm RS}(\theta)e_{t, \rm RS}(\theta)}{\sum_{\rm RS}w_{\rm RS}(\theta)}.
\end{equation}

The random catalog is constructed by distributing points uniformly within the survey's footprint, ensuring they respect the same angular mask and selection function as the lens sample. 

We created 6 random catalogs, one for every \maglim\ tomographic bin, ensuring the size of each random catalog to be exactly 50 times the size of the lens catalog for that specific tomographic bin. We verified the adequacy of the randoms catalog size in Section \ref{sec:impact}.

% We verified that the measurement is stable against the particular realization of the random catalog by repeating the analysis with an independent set of randoms of equal size ($50\times$ the lens sample). The resulting data vectors differ by $\Delta \chi^2 \simeq 5$, reported in Table \ref{tab:res_chi2} and shown in Figure \ref{fig:residuals_corrections}. This level of discrepancy is consistent with the stochastic noise expected from finite random sampling and numerical precision differences between runs. We therefore conclude that the adopted random catalog size is sufficient and that the measurement is robust to the choice of random realization. 

\subsection{Shear response}\label{sec:response}

%The source catalog used in Y6 is obtained using METADETECT, an upgrade of the Y3 METACALIBRATION software. They share the main feature, which is to be able to self-calibrate shear. What they do in practise, is to apply different amounts of shear to a galaxy, and calculate how the shear estimator responds to the applied shear. The Y6 upgrade included the detection step in the calibration, therefore the shear is applied to a whole image and not to single galaxies. The galaxy ellipticity is what we actually measure, and we can relate it to the shear through the response R. Since shear has two components, R is a 2x2 matrix, and when projecting R to the tangential component, we should perform a rotation of R. We are not doing that, and we have to justify that this is a sufficiently good approximation. Also, we are using a single R estimate per source tomographic bin, instead of using a value for each single angular bin considered in the gglensing measurement. We have to justify this approximation.

The source catalog used in Y6 is obtained using \textsc{Metadetection} \citep{Sheldon2023}, an upgrade of the Y3 \textsc{Metacalibration} software \citep{SheldonMcal2017}. They share the main feature, which is to be able to self-calibrate shear using the shear response factor, $\boldsymbol{R}$. The response is used to relate the two-component ellipticity to the two-component shear via the following Taylor expansion:
\begin{equation}
    \begin{aligned}
        \boldsymbol{e}(\boldsymbol{\gamma}) & \approx \boldsymbol{e}\vert_{\gamma=0} + \left.\frac{\partial \boldsymbol{e}}{\partial \boldsymbol{\gamma}}\right\vert_{\gamma=0}\boldsymbol{\gamma} + \mathcal{O}(\boldsymbol{\gamma}^2) \\
        & \equiv \boldsymbol{e}\vert_{\gamma=0} + \boldsymbol{R}\boldsymbol{\gamma} + \mathcal{O}(\boldsymbol{\gamma}^2),
    \end{aligned}
    \label{eTaylor}
\end{equation}
where $\boldsymbol{R}$ is defined as the first derivative of the ellipticity with respect to the shear. At first order, assuming that the intrinsic ellitpicities of galaxies are randomly oriented, we can therefore obtain the following relation by averaging over all ellipticities:
\begin{equation}
    \langle \boldsymbol{e} \rangle \approx \langle \boldsymbol{R \gamma} \rangle.
\end{equation}
We can invert this and then apply the tangential rotation from Eq. \ref{e_tangential} to give the average tangential shear:
\begin{equation}
    \langle \gamma_t \rangle \approx \langle \boldsymbol{R}_t \rangle^{-1} \langle e_t \rangle.
    \label{eq:R_t}
\end{equation}

In practise, we determine the response by applying an artificial shear to each galaxy, and measuring their sheared ellipticites. We can then calculate $\boldsymbol{R}$ using a finite-difference approximation:
\begin{equation}
    R_{ij} \approx \frac{e_i^+ - e_i^-}{\Delta\gamma_j},
    \label{eq:R_calculation}
\end{equation}
where $e_i^+$ and $e_i^-$ are the $i$ component of the ellipticities after applying a shear in the $j$ direction of $+\gamma_j$ and $-\gamma_j$, respectively. We use an artificial shear of $\gamma_j = 0.01$, so $\Delta\gamma_j = 0.02$. 

%In practise, this is done by applying an artificial shear to each galaxy, and then calculating how the shear estimator responds to the applied shear. With METADETECT, galaxy detection is included in the calibration process, so the artificial shear is applied to entire images rather than individual galaxies. 

\begin{table}
	\centering
	\caption{Mean values of the shear response in each tomographic source bin that are used in the computation of the tangential shear.}
	\label{tab:mean_responses}
	\begin{tabular}{cc}
        \\
		\hline
		  Source Bin & Mean Response\\
        \hline
	      1 & $0.857$\\
        2 & $0.868$\\
        3 & $0.820$\\
        4 & $0.692$\\
        \hline
	\end{tabular}
\end{table}

We make two approximations about the response matrix in order to simplify calculations. The first is the assumption that the matrix can be approximated as a scalar, and the second is that the response has no dependence on angular scale.
The response is therefore reduced to a single mean value for each tomographic source redshift bin, which are listed in Table \ref{tab:mean_responses}. We assess the validity of these approximations in Sec. \ref{sec:response_tests}.

%How do we actually calculate R?
%Average sheared ellipticities
%Tangential rotation required in theory?

\begin{figure}
    \includegraphics[width=0.45\textwidth]{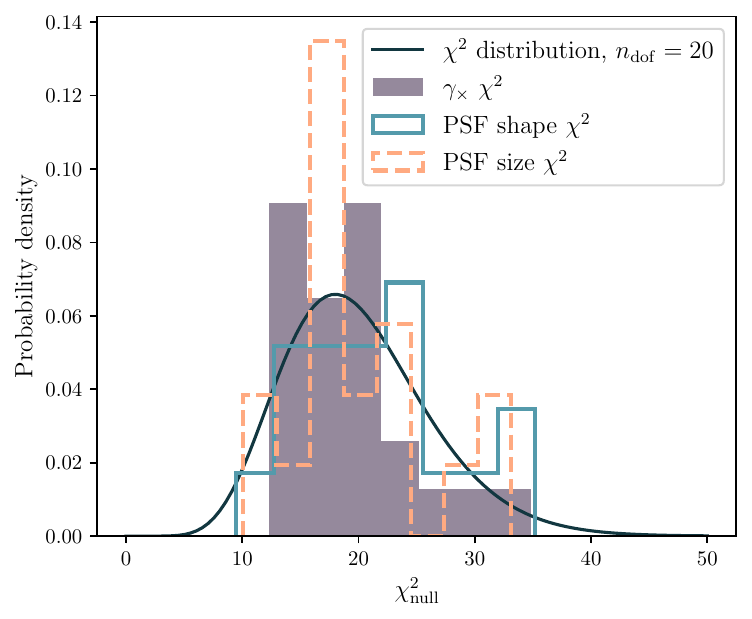}
    \caption{$\chi^2$ distributions for the null tests of $\gamma_\times$ (solid purple), the PSF shape-residual (teal), and the PSF size-residual (dashed salmon), evaluated across all lens-source bin combinations. The histograms show the measured $\chi^2$ values, while the black curve represents the theoretical $\chi^2$ distribution for 20 degrees of freedom (matching the 20 angular bins used). All tests are performed on the angular range $2.5' < \theta < 250'$. This comparison quantifies the consistency of residual signals with zero across all bin pairs.}
    \label{fig:chi2_both}
\end{figure}

% \begin{figure*}
%     \centering
%     \includegraphics[width=\textwidth]{Figures/random_points_250.pdf}
%     \caption{Tangential shear measured around random points, for each source tomographic bin. The shaded areas correspond to the scales where the signal might be affected by astrophysical systematics, which therefore are excluded from our analysis. [\textit{large scales to be removed in the plot}]}
%     \label{fig:random_points}
% \end{figure*}

\subsection{Final Corrected Estimator}\label{sec:final_estimator}

Our final estimator is then expressed as: 
\begin{equation}
    \gamma_{t} = \frac{1}{R} \left(\frac{\sum_{\rm r} w_{\rm r}}{\sum_{\rm l} w_{\rm l}} \frac{\sum_{\rm LS} w_{\rm LS}(\theta) e_{t, \rm LS}(\theta)}{\sum_{\rm RS}w_{\rm RS}(\theta)} - \frac{\sum_{\rm RS} w_{\rm RS}(\theta) e_{t, \rm RS}(\theta)}{\sum_{\rm RS}w_{\rm RS}(\theta)}\right).
    \label{eq:final_estimator}
\end{equation}

\begin{figure*}
    \includegraphics[width=0.49\linewidth]{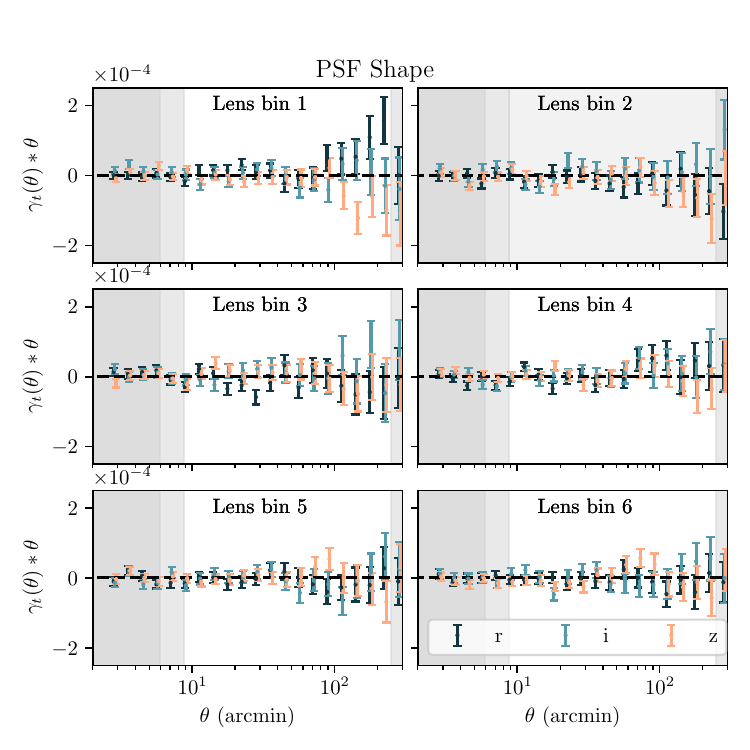}
    \includegraphics[width=0.49\linewidth]{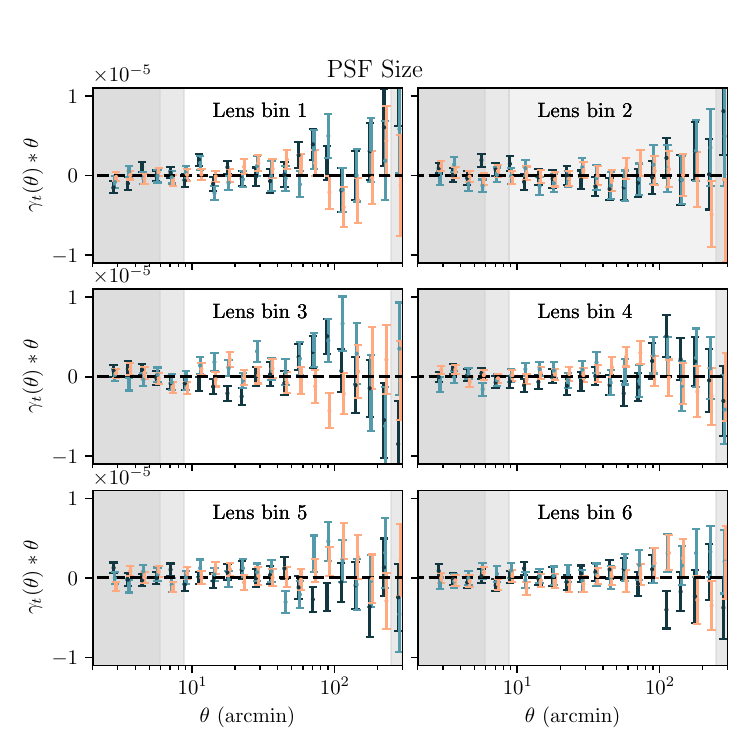}
    \caption{Left panel) Tangential shear of the shape residuals, defined as the difference between the measured star shapes and the corresponding PSF model shapes at their locations. The error bars are obtained from the Jackknife covariance. The results are shown for each combination of photometric band (r, i, z) and tomographic lens bin, demonstrating how the PSF model residuals translate into a weak-lensing signal. Right panel) Tangential shear measurements obtained after incorporating the size residuals of the PSF model. Size residuals are computed analogously to shape residuals, but using the stellar size differences instead. The star shape measurements are then scaled by these size residuals and the resulting tangential shear is shown for each photometric band (r, i, z) and tomographic lens bin. In both cases, Gray shaded regions mark the angular scales removed from the analysis: darker gray for the linear bias scale cuts and lighter gray for the nonlinear bias cuts. Lens bin 2 is fully shaded because it is excluded from the fiducial data vector. }

    \label{fig:psf_resids}
\end{figure*}

% \begin{figure}
%     \centering
%     \includegraphics[width=0.45\textwidth]{Figures/psf_chi2_250only.pdf}
%     \caption{$\chi^2$ distribution for the fit of $\gamma_t$ of the shape residuals and size residuals, across all bin combinations. The histogram represents the measured $\chi^2$ values, while the overlaid curve shows the expected theoretical $\chi^2$ distribution for 20 degrees of freedom (corresponding to the 20 angular bins). This comparison assesses the goodness of fit across the bin combinations.}
%     \label{fig:psf_chi2}
% \end{figure}

\begin{figure*}
    \centering
    \includegraphics[width=\textwidth]{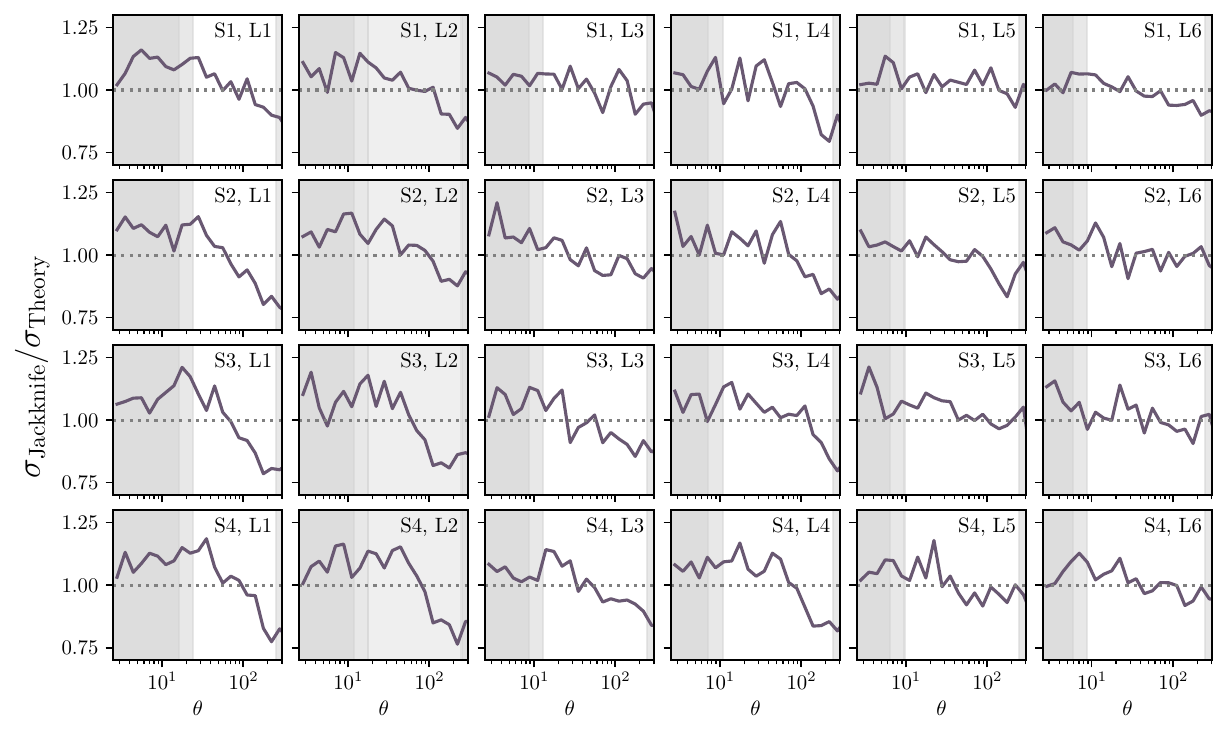}
    \caption{Ratio of the diagonal elements of the Jackknife covariance and the fiducial theoretical covariance. Gray shaded regions mark the angular scales removed from the analysis: darker gray for the linear bias scale cuts and lighter gray for the nonlinear bias cuts. Lens bin 2 is fully shaded because it is excluded from the fiducial data vector. Note that the theoretical covariance here does not account for point-mass marginalization, which is applied during likelihood evaluation through the inverse covariance matrix and therefore cannot be represented in terms of the covariance elements.}
    \label{fig:jackknifecov}
\end{figure*}

\begin{figure*}
    \centering
    \includegraphics[width=\textwidth]{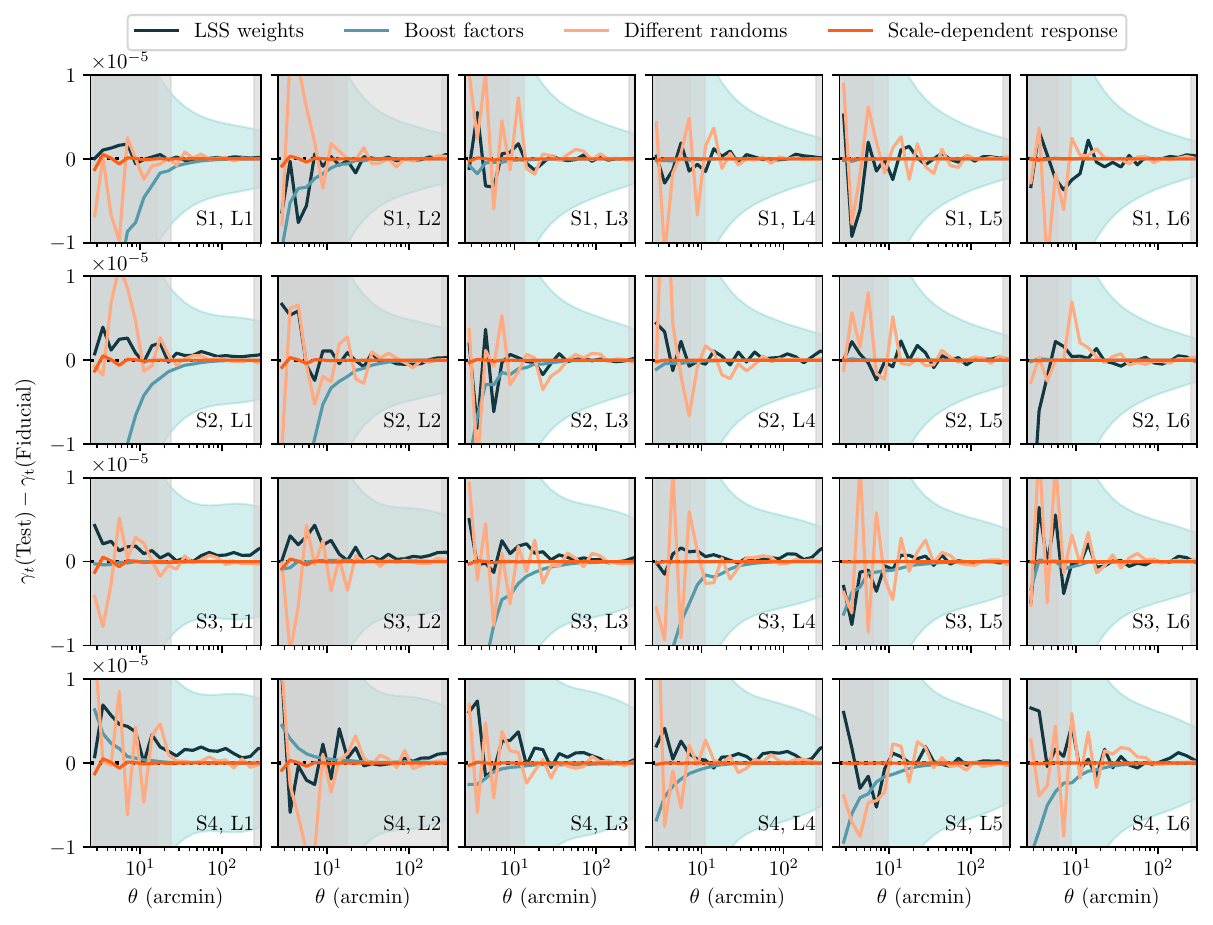}
    \caption{Residuals of galaxy-galaxy lensing measurements obtained by applying different configurations of large-scale structure (LSS) weights, boost factors, and random catalogs, shown with respect to the fiducial measurement. The transparent turquoise filled area represents the uncertainty estimated from the theory covariance. This comparison highlights the impact of these variations on the overall signal and the robustness of the analysis. The $\chi^2$ are reported in Table \ref{tab:res_chi2}. Gray shaded regions mark the angular scales removed from the analysis: darker gray for the linear bias scale cuts and lighter gray for the nonlinear bias cuts. Lens bin 2 is fully shaded because it is excluded from the fiducial data vector. Note that this figure does not illustrate the point-mass marginalization, which is applied during likelihood evaluation through the inverse covariance matrix and therefore cannot be represented in terms of the covariance elements shown here.}
    \label{fig:residuals_corrections}
\end{figure*}

In Figure~\ref{fig:gammat}, we show the tangential shear measurement, performed using the estimator described above. The error bars are derived from the theory covariance presented in \citetalias{y6-3x2pt}, and the best-fit model obtained after unblinding is overlaid. The total signal-to-noise ratio (S/N) is 173 when considering all scales and tomographic bins, and 75 after applying the scale cuts and excluding bin 2.

\section{Measurement: robustness tests}
\label{sec:validation}

In this Section we will be performing the necessary tests to ensure the robustness of the tangential shear measurement performed using the final estimator written in the previous section. 

We note that except for the tests discussed in Section \ref{sec:impact}, all uncertainty estimates in this work rely on the jackknife covariance (see Section \ref{sec:jackknifecov}). In addition to providing a model-independent estimate of the covariance, the jackknife method captures survey-specific effects ensuring that these diagnostic tests are evaluated using the noise properties of the data itself. This makes the jackknife covariance particularly well suited for assessing internal consistency and identifying potential systematics.

While the tangential shear measurement itself is robust across all bins, the $3\times2$pt analysis revealed that bin 2 exhibited a significant deviation in the posterior of its first redshift mode. As no clear explanation was found after extensive dedicated checks, the bin was conservatively excluded from the fiducial analysis. We refer the reader to \citetalias{y6-3x2pt} for a full explanation of this decision.

\subsection{Null test: Cross-component}

As we have seen, gravitational lensing distorts background galaxy shapes in a coherent, radial pattern around lensing structures, producing a lensing effect in the tangential direction. Therefore the cross-component of shear, $\gamma_\times$, which is the shear component at 45 degrees with respect to the tangential direction, is expected to be zero in the absence of systematic uncertainties. Any non-zero values of $\gamma_\times$ could indicate the presence of systematic errors in the measurement process, such as imperfect point-spread function (PSF) correction, or errors in the galaxy shape measurements. Hence, $\gamma_\times$ is used as a null test to verify that the measured shear is due to gravitational lensing and is not biased by other sources.

The measurement of the cross component, $\gamma_\times$, is shown in Figure~\ref{fig:gammax} and is consistent with zero within the measurement uncertainties across all bin combinations. We quantify this consistency by computing a $\chi^2$ statistic over the full set of $\gamma_\times$ measurements. Figure \ref{fig:chi2_both} shows the distribution of $\chi^2$ for all bins combinations (in purple), compared with the $\chi^2$ distribution for the corresponding number of degrees of freedom ($n_{\text{dof}}$, i.e. the number of angular bins considered). The measured $\chi^2$ values fall within the expected range, suggesting that $\gamma_\times$ is consistent with zero. 
% In contrast, if the measured $\chi^2$ values had exceeded the expected range significantly, it may have indicated the presence of systematic errors or unexpected astrophysical effects that need to be addressed.

% \subsection{Null test: Tangential shear around random points}\label{sec:randoms}

% To ensure that the measured galaxy-galaxy lensing signal is actually caused by the underlying matter distribution probed by the lens galaxies lensing the background source galaxies, rather than arising from systematic errors or observational biases, we measure the tangential shear of the source galaxies around a catalog of random points and check its compatibility with a null signal. 

% In Figure \ref{fig:random_points} we can see the tangential shear of each individual source tomographic bin around a set of random points. We used the random catalog corresponding to the first lens tomographic bin. The agreement with zero is good, quantified by the reduced $\chi^2$ (with applied scale cuts: 0.696, 0.528, 0.441, 0.357, for the four source tomographic bins respectively). The $\chi^2$ values are computed using the jackknife covariance  of the random point measurement. The slightly low reduced $\chi^2$ values therefore indicate that the jackknife errors may be somewhat conservative, or that correlations between data points reduce the effective number of degrees of freedom.

% P-values (0.778 0.880 0.923 0.967).

\subsection{Null test: PSF residuals}

%The Point Spread Function quantifies the blurring effect caused by the telescope optics, atmosphere, and other instrumental factors on the observed images of galaxies. In order to validate the PSF model, we measure the tangential shear of a catalog from GAIA stars around lens galaxies, after subtracting the PSF model computed for each of the stars. We expect the signal to be compatible with 0. I have performed this test dividing the lens galaxies by tomographic bin, but since the PSF model is derived for each photometric band, we should also check that we get results consistent with 0 for g, r, i, z

The Point Spread Function (PSF) quantifies the blurring effect caused by the telescope optics, atmosphere, and other instrumental factors on the observed images of galaxies. In order to validate effects of the PSF model \citepalias{y6-piff} on the measured tangential shear, we perform two null tests using a catalog of known GAIA stars that appear within the DES data sample. 

In the first test we consider the effect of the PSF model shape. We calculate the shape residuals, the differences between the measured shape of the stars and the PSF model shape at the same locations. Then we compute the tangential shear of the residuals around the lens galaxies for each combination of photometric band (r, i, z) and tomographic lens bin, which is shown in the left-hand panel of Figure \ref{fig:psf_resids}. 

The second test validates the effect of the PSF model size. We calculate the size residuals in the same way as the shape residuals, now using the star size and model size instead of the shape. The shape measurements for each star are scaled by the size residuals, and then used to calculate the tangential shear for each band and tomographic lens bin. We show the results in the right-hand panel of Figure \ref{fig:psf_resids}.

For both shape and size residuals, we find that the measured tangential shear is consistent with the null hypothesis in all three bands, both before and after applying scale cuts. In Figure \ref{fig:chi2_both} we show the $\chi_\text{null}^2$ distributions for shape and size (respectively in teal and salmon), each containing all combinations of photometric band and tomographic lens bin for all 20 data points. The expected $\chi_\text{null}^2$ distribution is also plotted for comparison.

\subsection{Covariance comparison}

The estimation of uncertainties on the measured tangential shear signal relies on an accurate characterization of the covariance matrix. Throughout this analysis, we adopt as our fiducial choice the theoretical covariance derived in \citetalias{y6-3x2pt}, computed from analytical predictions of the Gaussian and non-Gaussian components of the shear field, including shape noise and sample variance contributions. After unblinding, the covariance was updated to reflect the final survey characteristics--such as the unblinded redshift distributions, masks, and source noise properties--as well as the cosmology used in its computation, while maintaining the same analytical framework. This approach provides a smooth and noise-free estimate of the expected statistical uncertainty under a given cosmological model and survey geometry, as well as sample variance across the footprint, and is therefore particularly suitable for parameter inference.

While theoretically motivated, such an approach inherently relies on assumptions about the underlying cosmology, survey window function, and noise properties. For this reason, it is essential to validate the theoretical covariance against an empirical estimate derived directly from the data. As introduced in Section \ref{sec:final_estimator}, we use the \textit{jackknife} covariance, which is computed by dividing the survey footprint into $N_{\mathrm{JK}}$ approximately equal-area regions and repeatedly recalculating the measurement while omitting one region at a time. The covariance is then obtained from the variance among these pseudo-independent realizations. This resampling technique, implemented through \textsc{TreeCorr} \citep{treecorr}, captures both shape noise and survey geometry effects without relying on external simulations.

Figure~\ref{fig:jackknifecov} compares the theoretical and jackknife covariance matrices for our fiducial measurement. The level of agreement between the two estimates is good across all tomographic bins and scales, indicating that the theoretical covariance provides an accurate description of the statistical uncertainties in the data. Small differences at the largest angular scales are consistent with noise in the jackknife estimate, reflecting the finite number of regions and partial independence between them. Compared to Y3, the agreement is improved (the difference reported was $\sim 10^{-4}-10^{-5}$, while in Y6 is $\sim 10^{-6}$. Overall, this comparison validates the use of the theoretical covariance as our baseline choice for cosmological inference.

\subsection{Response approximations}
\label{sec:response_tests}

\subsubsection{Scalar approximation}

\begin{table}
	\centering
	\caption{Mean values of the response matrix computed from all source galaxies. The diagonal elements differ by just $0.2\%$, and the non-diagonal elements are negligible in comparison.}
	\label{tab:response_matrix}
	\begin{tabular}{cc}
        \\
		\hline
		  Matrix Component & All Bins\\
        \hline
	      $R_{11}$ & $0.8178$\\
        $R_{12}$ & $0.0012$\\
        $R_{21}$ & $-0.0070$\\
        $R_{22}$ & $0.8193$\\
        \hline
	\end{tabular}
\end{table}

The first of our two assumptions regarding the shear response matrix is that the response is independent of the relative orientation of galaxies. This means that a tangential rotation does not need to be applied to the matrix (see Eq. \ref{eq:R_t}), and that the matrix can hence be treated as a single scalar value. To test that this approximation is valid, we compute the full $2 \times 2$ response matrix. As the response is a noisy measurement, we do this using the average artificially-sheared ellipticites of the full sample to minimise the impact of noise. The results, listed in Table \ref{tab:response_matrix}, show that the diagonal elements of the matrix, $R_{11}$ and $R_{22}$, differ by just $0.2\%$. This is consistent with previous results from Y3, where the differences for individual tomographic bins were calculated to be between $0.01\%$ and $0.1\%$. The non-diagonal elements, $R_{12}$ and $R_{21}$, are $\sim100$ times smaller than the diagonal elements. We therefore conclude that it is valid to approximate the response as a scalar, calculated as: 
\begin{equation}
    R \approx \frac{R_{11} + R_{22}}{2}.
    \label{scalar_R}
\end{equation}

% \begin{table*}
% 	\centering
% 	\caption{Effect of corrections to the tangential shear estimator. The impact is quantified by $\Delta \chi ^2$ values, for the full scale range (624 data points), for the scales considered after scale cuts (312 data points, see Section \ref{sec:scalecuts}), after removing lens bin 2 (264 data points), and using point-mass marginalisation (264 data points, see Section \ref{par:pm}). Values in brackets were computed using nonlinear scale cuts.}
% 	\label{tab:correction_impacts}
% 	\begin{tabular}{ccccc}
%         \\
% 		\hline
% 		   Correction & $\Delta \chi^2$ (all scales) & $\Delta \chi^2$ (scale cuts) & $\Delta \chi^2$ (scale cuts, no bin2) & $\Delta \chi^2$ (scale cuts, no bin2, PM)\\
           
%         \hline
% 	      LSS weights & 9.21 & 2.61 & 2.18 & 2.50\\
%           Boost factors & 10.51 & 2.14 & 1.77 & 1.78\\
%           Diff. randoms & 14.43 & 6.36 & 4.97 & 5.06\\ 
%           Scale-dep. Response & 0.14 & 0.0046 (0.0052) & 0.0031 (0.0038) & 0.0032 (0.0040) \\
%         \hline
% 	\end{tabular}\label{tab:res_chi2}
% \end{table*}

\begin{table*}
    \centering
    \caption{Effect of corrections to the tangential shear estimator. The impact is quantified by $\Delta \chi ^2$ values, for the full scale range (624 data points), for the scales considered after scale cuts (shortened as \textit{sc}, 312 data points, see Section \ref{sec:scalecuts}), after removing lens bin 2 (264 data points), and using point-mass marginalisation (see Section \ref{par:pm}).}
    \label{tab:res_chi2}

    \begin{tabular}{c c cc cc cc}
        \hline
        & $\Delta \chi^2$ (all scales) 
        & \multicolumn{2}{c}{$\Delta \chi^2$ (\textit{sc})} 
        & \multicolumn{2}{c}{$\Delta \chi^2$ (\textit{sc}, no bin 2)} 
        & \multicolumn{2}{c}{$\Delta \chi^2$ (\textit{sc}, no bin 2, PM)} \\
        \cline{3-8}
        Effect 
        &  
        & Linear & Nonlinear 
        & Linear & Nonlinear 
        & Linear & Nonlinear \\
        \hline

        LSS weights 
        & 9.21 
        & 2.61 & 2.95 
        & 2.18 & 2.48 
        & 2.15 & 2.37 \\

        Boost factors 
        & 10.51
        & 2.14 & 2.64 
        & 1.77 & 2.18 
        & 1.73 & 2.01 \\

        Diff. randoms 
        & 14.43
        & 6.36 & 7.72 
        & 4.97 & 6.37 
        & 5.88 & 6.32 \\

        Scale-dep. Response 
        & 0.14 
        & 0.0046 & 0.0052 
        & 0.0031 & 0.0038 
        & 0.0032 & 0.0040 \\
        \hline
    \end{tabular}

\end{table*}

\subsubsection{Scale-dependence approximation}

We also make the assumption that the response has no dependence on angular scale. This means that we can average the response over an entire tomographic source bin, instead of having to calculate a different average for each angular bin. We test the impact of this approximation by computing the tangential shear both with and without the approximation, and then calculating the residuals between the two datavectors. 

In order to determine the scale-dependent responses, we compute the scalar correlation functions between the lens and source galaxies, in which the artificially-sheared ellipticities used in Eq. \ref{eq:R_calculation} are treated as scalar quantities associated with the source galaxies. This gives the scale-dependent averages of $e_i^+$ and $e_i^-$ for $i = 1, 2$, which can then be used to calculate the scale-dependent responses by applying Eq. \ref{eq:R_calculation} and Eq. \ref{scalar_R}. We are then able to use these responses in our tangential shear estimator (Eq. \ref{eq:final_estimator}) in place of the full tomographic bin averages used in the fiducial analysis. 

We evaluate the impact of this approximation by comparing the measurement obtained using scale-dependent responses to the fiducial measurement that adopts a single response per tomographic bin. The resulting residuals and their statistical significance are presented in the next Section~\ref{sec:impact}, where the impact of all estimator corrections is assessed jointly. This subsection focuses solely on the construction of the test.

% We show the residuals between the test case (using the average response per angular bin) and the fiducial case (using the average response per redshift bin) in Fig. \ref{fig:residuals_corrections}. At all angular scales, the residuals are within the uncertainties on the fiducial measurement. We show the $\Delta\chi^2$ between the test and fiducial measurements in Table \ref{tab:correction_impacts}, with values ranging from 0.0022 to 0.1405 depending on the choice of scale cuts and covariance. The approximation therefore has a negligible impact on the tangential shear and is valid to make. 

%In Figure \ref{fig:residuals_corrections} the impact of using a scale-dependent response correction is shown, with respect to the fiducial approach. 

% \subsection{Impact of effects on tangential shear measurement}

\subsection{Impact of effects}\label{sec:impact}

Here we quantify the impact of four distinct components of the tangential-shear estimator. In particular, we repeat the measurement under the following modifications:
\begin{itemize}
\item omitting the Large-Scale Structure (LSS) weights applied to the \maglimpp\ lens sample, which mitigate spatial variations in galaxy density (Section~\ref{sec:data});
\item omitting the boost-factor correction, which accounts for physically associated lens-source pairs (Section~\ref{sec:boosts});
\item replacing the fiducial random catalog with an independent realization of equal size when computing the random-lens and random-source pair counts entering Eq.~\ref{eq:final_estimator};
\item adopting a scale-dependent shear response in place of the tomographic-bin-averaged response used in the fiducial pipeline (Section~\ref{sec:response_tests}).
\end{itemize}

For each test, we subtract the fiducial tangential-shear data vector and show the resulting residuals in Figure~\ref{fig:residuals_corrections}. The corresponding $\Delta\chi^2$ values, computed using the fiducial theoretical covariance, are reported in Table~\ref{tab:res_chi2}. We evaluate four summary statistics for each test: using the full scale range; after applying the fiducial scale cuts; after removing lens bin 2 (whose inclusion results in internal tension in the $3\times2$ data vector); and after marginalizing over the point-mass term (Section~\ref{par:pm}).

Omitting the LSS weights or the boost factors results in modest deviations from the fiducial measurement, with $\Delta\chi^2 \simeq 1-2$ after scale cuts. These differences are small relative to the statistical uncertainties and indicate that the corrections play a subdominant but non-negligible role in stabilizing the estimator on small and intermediate scales.

% To test sensitivity to the particular realization of the random catalog, we recomputed the measurement using a fully independent set of random points. The resulting discrepancy ($\Delta\chi^2 \simeq 6$) is consistent with the level of stochastic variation expected from finite random sampling and numerical precision, and does not indicate any systematic bias in the estimator (see Section~\ref{sec:randoms}). 
To test sensitivity to the particular realization of the random catalog, we repeated the measurement using an independent set of random points with the same construction and size (see Section~\ref{sec:randoms}). The resulting difference corresponds to $\Delta\chi^2 \simeq 6$, which quantifies the change between the two measurements in data space. This test probes the stability of the estimator with respect to random-catalog realizations, rather than a physical systematic in the signal model. While increasing the random catalog size suppresses Poisson noise in the random counts, differences between random realizations can still enter coherently through survey geometry and weighting, and can therefore lead to a residual $\Delta\chi^2$.
This confirms that the adopted random catalog size is sufficient for a stable measurement.

Finally, adopting a scale-dependent shear response instead of a tomographic-bin-averaged one produces residuals that are orders of magnitude smaller than the statistical uncertainties, with $\Delta\chi^2 < 0.01$ across all cases considered. This validates the approximation used in the fiducial pipeline (Section~\ref{sec:response_tests}).

Overall, all four tests demonstrate that the impact of these corrections on the galaxy-galaxy lensing signal is small, and that the fiducial measurement is robust within the statistical precision of the DES~Y6 dataset.

\section{The shear ratio test}
\label{sec:shearratio}

The shear-ratio is defined as the ratio of two tangential shear measurements from different source bins but using the same lens. This combination cancels exactly the dependence on the matter around the lens (including the galaxy-matter power spectrum and galaxy bias) in the idealized limit of an infinitely narrow lens sample, and with no overlap between the lens and source redshift distributions, yielding a residual signal only sensitive to the angular diameter distances of the source and lens populations.  

Historically, the shear-ratio test was devised to constrain cosmological parameters \citep{Jain_2003},  especially dark energy's evolution in time. However, due to its sensitivity to uncertainties in the source redshift distribution ($z_s$) and to multiplicative shear calibration biases, as well as to intrinsic alignments in cases where the lens and source samples overlap, it has more recently been used as an internal validation and calibration test for these systematic effects (e.g., \citep{Blazek_2012}, \citep{y3-shearratio}, \citep{emas_2025},  \citep{hsc3_sr}). It is particularly sensitive to faint, high-redshift galaxies, which are the most susceptible to systematic uncertainties but provide the largest GGL signal (see Figure~\ref{fig:gammat}). A robust calibration of the redshift distribution is important, particularly because errors in the mean redshift can significantly impact the cosmological constraints. For this analysis, we employ the geometrical shear-ratios to verify whether the SR test is satisfied.

\subsection{Modeling of the ratios}

\subsubsection{Geometrical Model}
\label{sec:geom_model}
In previous studies, such as \citep{y1-gglensing} and \citep{hsc3_sr}, the shear-ratio test has been applied in the geometrical framework. In the present analysis, we adopt the same approach, applying it within the geometrical regime. This approach stems from the definition of the tangential shear, which is given by the ratio of the excess surface mass density of the lens to the critical surface mass density: $\gamma_t = \Delta \Sigma / \Sigma_{\text{crit}}$. By taking the ratio using the same foreground lens, the numerator cancels out, simplifying the measurement to a function of the commoving angular diameter distances:

\begin{equation}
    \frac{\gamma_t^{{\rm l}_i,{\rm s}_j}}{\gamma_t^{{\rm l}_i,{\rm s}_k}}=\frac{\Sigma_\text{crit}^{-1} (z_{{\rm l}_i}, z_{{\rm s}_j})}{\Sigma_\text{crit}^{-1}(z_{{\rm l}_i}, z_{{\rm s}_k})}, \quad \text{where:} \quad \Sigma_\text{crit}^{-1} (z_{\rm l},z_{\rm s}) = \frac{4\pi G}{c^2} \frac{D_{\rm LS}D_{\rm l}}{D_{\rm s}};
    \label{eq:ratiogt}
\end{equation}

with $\Sigma_{\mathrm{crit}}^{-1}(z_{\rm l}, z_{\rm s}) = 0$ for $z_{\rm s}<z_{\rm l}$,  and where $z_l$ and $z_{\rm s}$ are the lens and source galaxy redshifts, respectively. Given that the redshift distributions we work with have a non-negligible width and may overlap between bins, we have to integrate over the full redshift distribution for each lens bin (i) and source bin (j):
\begin{equation}
    \Sigma_{\text{crit,eff}}^{-1 \, i,j} = \int_0^{z_{\rm l}^{\text{max}}} dz_{\rm l} \int_0^{z_{\rm s}^{\text{max}}} dz_{\rm s} \, n_{\rm l}^i(z_{\rm l}) \, n_{\rm s}^j(z_{\rm s}) \, \Sigma_{\text{crit}}^{-1}(z_{\rm l}, z_{\rm s})
\end{equation}
Therefore, the geometrical ratio depends on the redshift distributions of the source and lens populations, the underlying cosmological parameters through their dependence on the angular diameter distances, and also the multiplicative shear bias. When using this approach, we fix the cosmology to $\Omega_m = 0.31$ and $H_0 = 67 \text{ km s}^{-1} \text{Mpc}^{-1}$. As we will explain in the next sections, the dependence on the choice of cosmological parameters is minimal.

In addition, the multiplicative shear bias ($m_j$) must be accounted for, as this parameter is expected to affect the galaxy-galaxy lensing signal (see Eq. \ref{eq:addm}). This effect is introduced as an independent multiplicative parameter in Eq. ~\eqref{eq:ratiogt} for each source redshift bin
\begin{equation}
    \frac{\gamma_t^{{\rm l}_i,{\rm s}_j}}{\gamma_t^{{\rm l}_i,{\rm s}_k}}\simeq\frac{(1+m_j)}{(1+m_k)}{\frac{\Sigma_\text{crit,eff}^{-1\;{\rm l}_i,{\rm s}_j}}{\Sigma_\text{crit,eff}^{-1\;{\rm l}_i,{\rm s}_k}}}
    \label{eq:sr_geom}
\end{equation}

Moreover, in this analysis we will apply two different models to calibrate the redshift uncertainties, to check whether the SR is satisfied. The fiducial Y6 redshift calibration, based on the mode projection method described in Section~\ref{sec:modelling}, and the shift method, which was employed in previous DES analyses. The latter approach models the redshift uncertainties by shifting the mean of the redshift distributions $n^i(z) = \bar{n}^i(z - \Delta z^i)$, where $\bar{n}^i(z)$ denotes the mean redshift distribution for bin $i$. Meanwhile, the fix values of the multiplicative shear bias, $m_j$, are taken from the blending corrections presented in \citetalias{y6-imagesims}.

\subsection{Measurements of the ratios}

For the measurement of the shear ratio, we compute the ratio between $\gamma_t^{\,{\rm l}_i, {\rm s}_j}(\theta)$ and $\gamma_t^{\,{\rm l}_i, {\rm s}_k}(\theta)$, which corresponds to using the same lens sample ${\rm l}_i$ but different source samples ${\rm s}_j$, ${\rm s}_k$. For simplicity, and since the ratios are predominantly geometrical and thus largely scale-independent \citep{y3-shearratio}\citep{Emas_2024}, we average the measured SR across the different angular bins. 

\begin{equation}
r^{({\rm l}_i, {\rm s}_j, {\rm s}_k)} \equiv \left\langle \frac{\gamma_t^{\,{\rm l}_i, {\rm s}_j}(\theta)}{\gamma_t^{\,{\rm l}_i, {\rm s}_k}(\theta)} \right\rangle_{\theta} =  \left\langle  r^{({\rm l}_i, {\rm s}_j, {\rm s}_k)}(\theta) \right\rangle_{\theta},
\label{r_mean}
\end{equation}

where the average over angular scales, includes the corresponding correlations between measurements at different angular scales \citep{y3-shearratio}. A specific combination is described by $({\rm l}_i, {\rm s}_j, {\rm s}_k)$, where ${\rm l}_i$ corresponds to the lens bin, and ${\rm s}_j$ and ${\rm s}_k$ the different source redshift bin, where ${\rm s}_k$ specifies the source bin used for the denominator of the ratio. To obtain the covariance of the shear-ratios, we use the theoretical covariance of the galaxy--galaxy lensing measurements described in Section \ref{sec:measurement}. We generate $10^5$ realizations of the GGL, compute the shear ratio for each realization, and then average over the angular scales \citep{y3-shearratio}. These are then used to estimate the covariance between the different shear ratios.

To ensure statistical independence from the galaxy-galaxy lensing 2pt likelihood analysis, we restrict the SR to small-scale information. The angular separation is bounded by a minimum value of $\theta_{\min} = 2.5 \text{ arcmin}$, which corresponds to the smallest scale at which the tangential shear is validated in this work. The upper boundary is set by a maximum of $6 h^{-1} \text{ Mpc}$. This choice is implemented to align with the small-scale cut-off used for the galaxy-galaxy lensing analysis. Since point-mass marginalization is used in the 3$\times$2pt analysis as described in Sec.~\ref{par:pm}, this ensures that any dependency of the tangential shear on scales smaller than the scale cut of $6 h^{-1}$ Mpc is removed. Consequently, the SR measurement will not be correlated with both the cosmic shear and the large-scale components of the  3$\times$2pt data vector, making it an independent probe.

For the geometrical combinations, we use the lens sample from \textsc{MagLim} Y6 \citepalias{y6-maglim}, and the source samples derived from \textsc{Metadetection} \citepalias{y6-metadetect} and image simulations \citepalias{y6-imagesims}. While \textsc{Metadetection} self-calibrates the shear measurements from the data itself, we still assess residual biases using  realistic image simulations which we parametrize with the multiplicative shear bias parameters. Therefore, when using information from image simulations, the bias in the multiplicative shear calibration must be accounted for, as it directly impacts the inferred shear-ratios.

\subsection{Validation}

\begin{figure}
    \centering
    \includegraphics[width=0.48\textwidth]{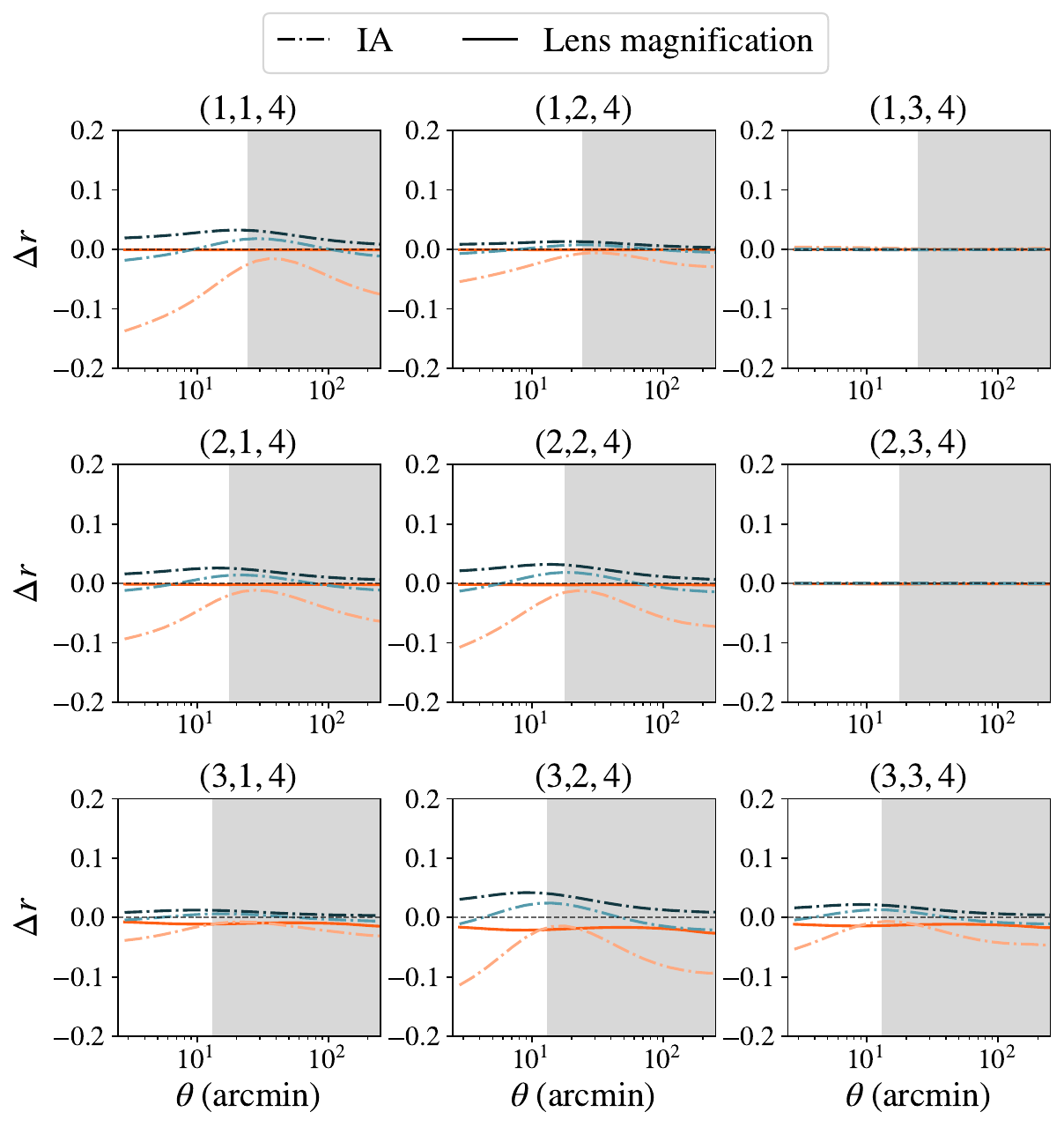}
    \caption{ Shear-ratio full model as a function of all scales, representing the effect of intrinsic alignment and lens magnification parameters. We find that for two redshift combinations, (1,3,4) and (2,3,4), the model is fully geometric. For the remaining combinations, we find that lens magnification can be neglected for lens bins 1 and 2, while it produces small changes for redshift bin 3. Intrinsic alignments, however, are significant in all cases (except the geometric ones). The dash-dot lines correspond to different IA TATT parameters within the following ranges: $\text{A}_1 = [-0.5,0.5],\text{A}_2 = [-1,2], \eta_1 = [-3,3],\eta_2 = [-3,3]$ with fixed $b_{TA} = 1$. For the lens magnification, we adopt $\alpha = [1.58,1.38,2.04]$, corresponding to the lowest three lens bins. The unshaded regions correspond to the angular scales we use in this analysis.}
    \label{fig:sr_IA_noIA}
\end{figure}

The geometrical modeling of the shear-ratio is only valid within certain regimes and does not capture the full range of effects that can influence the measurements. Therefore, we use the full modeling framework of \cite{y3-shearratio}, which incorporates all relevant astrophysical effects, to identify the combinations of redshift bins whose behavior is closest to the geometric prediction. These combinations are then selected for our Y6 shear-ratio analysis. Furthermore, we use this model for convenience as the fiducial one for running SR-only chains and for joint analyses in combination with other probes, such as cosmic shear and  3$\times$2pt. 

The full model includes integration over the matter power spectrum and contributions from systematics such as intrinsic alignments and lens magnification. It uses the tangential shear definition of Eq. \ref{eq:gammat_definition} to compute the ratio as a function of angular scale. The shear signal entering the ratio is modeled using the same framework adopted for the galaxy-galaxy lensing analysis described in Section~\ref{sec:modelling}, and the resulting ratio is averaged over angular scales in the same form used for the measurements (see Eq. \ref{r_mean}).

The validation of the full shear-ratio model was extensively performed in the DES Y3 analysis \citep{y3-shearratio}, testing a wide range of potential systematic effects. They showed that baryonic feedback, nonlinear galaxy bias, HOD modeling and boost factors have a negligible impact on the SR and on the derived parameter constraints, demonstrating that the adopted modeling framework is robust. In this paper, we adopted the same model, relying on the validation presented in DES Y3. To identify the redshift combinations that are “fully geometric”, we compare the full model with the geometric model. This means selecting the combinations of lensing ratios that are affected almost exclusively by multiplicative shear bias and source redshift distributions.

Other systematics effects that can impact the measurement of SR are intrinsic alignments (IA) and lensing magnification. In Figure ~\ref{fig:sr_IA_noIA}, we show the impact of different IA TATT models, as well as the fiducial lensing magnification values, in comparison to the scenario without these effects. As shown in Figure ~\ref{fig:sr_IA_noIA}, the lensing ratios (1,3,4) and (2,3,4) are not affected by IA and lensing magnification. Moreover, SR can be influenced by the choice of cosmological parameters. To assess this, we ran SR-only chains while fixing different cosmological parameters. We found that their effect was negligible.

We decide to restrict our analysis only using these combinations, which we will refer to as 'geometrical' SR. This choice ensures that the constraints derived from the shear ratios arise from their sensitivity to the source redshift distributions and the multiplicative shear bias, rather than from other nuisance parameters.

\subsection{Results}

We take the lensing ratios by selecting combinations that are fully geometrical, corresponding to the two highest source redshift distributions and the two lowest lens redshift bins. These choices ensure that our constraints are primarily sensitive to the source redshift uncertainties and the multiplicative shear bias, while being unaffected by other potential systematic effects. Therefore, any constraints on cosmological parameters can be attributed directly to the constraints on these two parameters.

\subsubsection{Consistency between shear-ratio and shear/redshift calibration}

\begin{figure}
    \centering
    \includegraphics[width=0.48\textwidth]{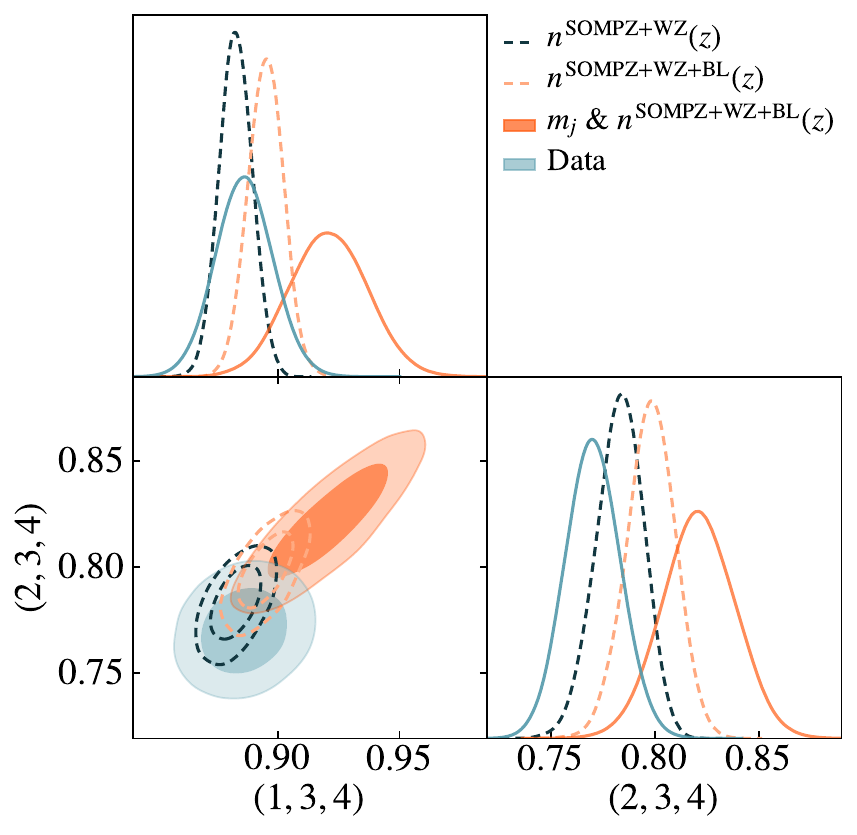}
    \caption{ Comparison of the SR prediction from the geometrical model and the SR measurements obtained from the angle-averaged ratios of the GGL combinations (Data) of the two geometrical shear ratios used in this analysis. We show the differences in the SR predictions coming from different redshift distributions $n(z)$, those derived from \textsc{SOMPZ+WZ} and \textsc{SOMPZ+WZ+BL}. The blending correction introduces a redshift dependence in $m_j$, which must be incorporated into the SR prediction.}
    \label{fig:sr_contour}
\end{figure}

To assess whether the shear-ratio test is satisfied, we check the consistency between the measured lensing ratios and the shear and redshift calibrations. 

We first compare the measured shear ratios with the geometrical predictions obtained using the fiducial redshift calibrations (see Figure~\ref{fig:sr_contour}). To compute the shear-ratio variations according to the geometrical model defined in Section~\ref{sec:geom_model}, we apply the \textit{mode projection} methodology. We generate $10^4$ realizations for each lens and source bin using Eqs.~\eqref{eq:modes_l} and \eqref{eq:modes_s}, respectively, producing $10^4$ random mode amplitudes ($u_i$). For each realization, we then calculate the corresponding shear ratio. Moreover, because blending corrections derived from image simulations introduce a bias in the multiplicative shear parameters, we generate an additional $10^4$ realizations centered at $m_3, m_4 = [0.016, 0.0017]$ \citepalias{y6-imagesims}. The results, presented in Figure~\ref{fig:sr_contour}, show that the geometrical model is consistent with the measurements derived from the data. The lensing ratios obtained from the GGL values are within $2.5\sigma$ of the purely geometrical predictions computed using the fiducial $n(z)$ used in the 3$\times$2pt cosmological analysis.

\begin{figure}
    \centering
    \includegraphics[width=0.48\textwidth]{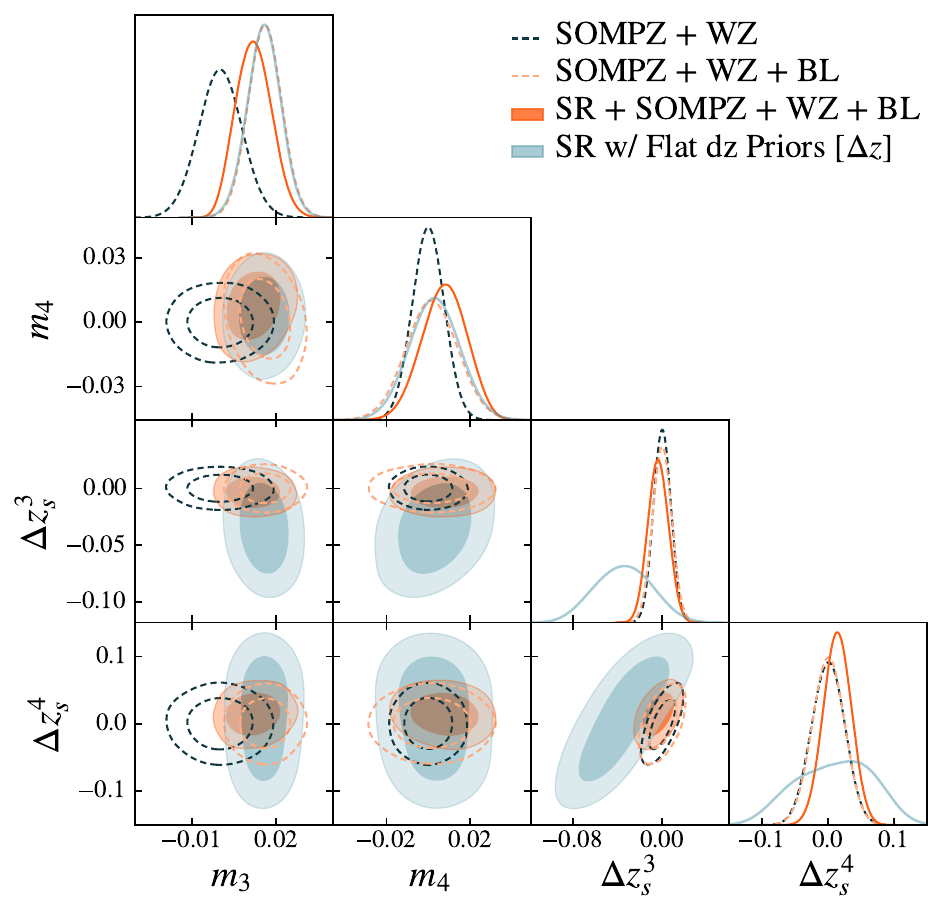}
    \caption{Mean source redshift and multiplicative shear bias constraints from the SR-only chains for the two highest redshift source bins, using the shift method, with flat priors $[-0.1,0.1]$ on $\Delta z_i$ only. This is compared with the combination of the mode projection redshift calibration method and its corresponding multiplicative shear calibration: \textsc{SOMPZ+WZ} and \textsc{SOMPZ+WZ+BL}. The SR information drives a shift toward lower $\Delta z_3$ values. When tight priors on $\Delta z_3-\Delta z_4$ from \textsc{SOMPZ+WZ+BL} are applied, the shear-ratio adds constrains on $m_3-m_4$.}
    \label{fig:sr_dz_m}
\end{figure}

We now perform the equivalent comparison in the model space, propagating the geometrical shear ratios to the source mean redshifts and the multiplicative shear bias parameters (see Figure \ref{fig:sr_dz_m}). We run SR-only chains employing two different redshift calibration methodologies: the mode projection and the shift method. To compare them better, we convert the modes information from the sources into shifts to the mean source redshifts, $\Delta z_s^j$. To achieve this, we generate $10^4$ variations using Eq. \ref{eq:modes_s} for the two highest source redshift bins. For the SOMPZ + WZ and SOMPZ + WZ + BL cases, the variations are drawn directly from its priors of the modes, while for SR + SOMPZ + WZ + BL, they are drawn from the posterior of the modes obtained from the SR-only MCMC. From these realizations, we then compute the corresponding shifts, $\Delta z_s^j$, of the mean of $n_j(z)$.

In Figure \ref{fig:sr_dz_m}, we present the SR constraints on the multiplicative shear bias and the source mean redshift for the two highest source bins, together with the redshift priors derived from \textsc{SOMPZ+WZ} and \textsc{SOMPZ+WZ+BL}. The priors on the $m_j$ parameters come from the \textsc{Metadetection} calibration and from the blending correction, respectively. Using this information, we test the consistency between the SR-only chain, which uses flat priors $\Delta z_s = [-0.1, 0.1]$, and the redshift priors. We find a difference of $1.68\sigma$ with the \textsc{SOMPZ+WZ} priors, and $1.59\sigma$ with the \textsc{SOMPZ+WZ+BL} priors, the latter being the fiducial redshift priors for the $3\times2\mathrm{pt}$ analysis. Since the SR-only constraints using shift method is independent of the \textsc{SOMPZ+WZ} and \textsc{SOMPZ+WZ+BL}, the agreement demonstrates the consistency of the fiducial redshift calibration and confirms that the SR test is successfully passed.

\section{Conclusions}

\label{sec:conclusions}

This work presents the final galaxy-galaxy lensing (GGL) measurement from the Dark Energy Survey (DES), based on the full six-year dataset. It builds on the methodologies established in previous data releases, particularly DES Y3, and represents the culmination of a decade-long effort to extract precise and robust lensing measurements from wide-field photometric data. With a signal-to-noise ratio of 173 over the full angular and tomographic range, this analysis yields the highest precision GGL measurement to date from any stage III survey. The increase in statistical power, combined with methodological advancements and improved control of systematics, marks a critical step in the transition from Stage III to Stage IV lensing experiments.

The measurement is based on the \textsc{MagLim++} lens sample and the \textsc{Metadetection} source catalog, which together span an effective area of $4031~\mathrm{deg}^2$ and yield an effective source number density of 8.29 galaxies per arcmin$^2$. Due to the deeper imaging available in Y6, the size of the source catalog increased by a factor of $\sim$1.5 with respect to the Y3 source catalog. The source catalog benefits from the application of \textsc{Metadetection}, which enables the self-calibration of selection and detection biases by applying artificial shears prior to object detection. This technique is essential for achieving accurate shear calibration at the depth of Y6 and significantly reduces the uncertainties on the multiplicative shear bias parameters. In contrast, the size of the \textsc{MagLim++} lens sample remained similar to Y3, reflecting a deliberate choice in the selection criteria. The lens sample selection incorporates additional color-color cuts and star-galaxy separation based on near-infrared imaging, as well as a SOM-based identification of regions of color space with high contamination, improving the purity of the sample without significantly degrading its statistical power.

Although six lens bins were originally defined, bin 2 was excluded from the fiducial data vector due to internal tensions identified in the $3\times2$pt analysis and unexplained deviations in its redshift calibration posterior \citepalias{y6-3x2pt}.

A defining feature of this analysis is the set of validation tests performed to ensure the robustness of the tangential shear measurement. We applied and tested all standard corrections, including boost factor corrections, random point subtraction, and shear response calibration, demonstrating that their impact on the final measurement is well within the expected statistical uncertainty after scale cuts. The combination of point-mass marginalization with physically motivated scale cuts (6 Mpc $h^{-1}$ for the linear bias model and 4 Mpc $h^{-1}$ for the nonlinear bias model) enables us to retain as much small-scale information as allowed by the robustness tests. These cuts exclude the scales where unmodeled small-scale contributions--such as higher-order galaxy-bias terms or baryonic effects--become non-negligible. We have also verified that the measurement is insensitive to the angular scale dependence of the shear response and to the non-diagonal elements of the shear response matrix. All null tests--including the cross-component of shear, the tangential shear around random points, and PSF residuals using Gaia stars--are consistent with expectations from noise and survey geometry, indicating the absence of significant additive or multiplicative systematics at the current level of precision.

In addition to the measurement itself, we perform a detailed comparison between the theoretical covariance matrix used in cosmological inference and a jackknife covariance derived from the data. The good agreement between the two validates the use of the analytic covariance, which benefits from lower noise and allows for accurate treatment of super-sample covariance and mode coupling. These results further demonstrate that the statistical uncertainties of this measurement are well understood, providing a solid foundation for subsequent parameter inference.

We also exploit the high signal-to-noise of the DES~Y6 GGL data to revisit the shear-ratio test as a consistency check on redshift and shear calibration. We select lens-source bin combinations where the tangential shear ratio is minimally affected by galaxy bias, intrinsic alignments, and magnification, such that the resulting observable is primarily sensitive to the source redshift distributions and the multiplicative shear bias. The measured ratios are consistent with the predictions derived from the fiducial redshift calibration based on \textsc{SOMPZ}, \textsc{WZ}, and image simulations, and constrain shifts in the mean redshift of the two highest source bins at the few percent level. These results provide an independent confirmation of the reliability of the redshift and shear calibration framework adopted for DES~Y6 cosmology analyses.

The galaxy-galaxy lensing signal presented here plays a central role in the DES~Y6 multi-probe program. As part of the 3$\times$2pt analysis, it enables simultaneous calibration of galaxy bias, photometric redshift uncertainties, and intrinsic alignments, while also contributing cosmological constraining power in its own right. 

In the broader context of cosmology, this work represents the final legacy GGL measurement from DES, and sets a new benchmark for wide-field photometric lensing surveys. The pipeline, methodology, and systematics control demonstrated here will inform future analyses from Stage IV surveys such as LSST, Euclid, and the Nancy Grace Roman Space Telescope. These next-generation experiments will probe the growth of structure and geometry of the Universe with higher precision, over wider and deeper datasets, and extending to smaller physical scales. The techniques validated in DES~Y6--particularly in handling redshift calibration, response corrections, and small-scale modeling--are expected to form the basis of those future analyses.

The increasing statistical precision of lensing measurements from current and future surveys demands a proportional improvement in our modeling and calibration of systematic effects. In that sense, this work illustrates both the current maturity of galaxy-galaxy lensing as a cosmological probe and the challenges ahead. While the DES~Y6 measurement is effectively free of detectable systematics at current sensitivity, further improvements in both theoretical modeling and calibration strategies will be necessary to fully exploit the potential of upcoming surveys. Nevertheless, the methodology and results presented here establish a reliable and robust foundation for the next generation of weak-lensing cosmology.

\section*{Acknowledgements}

{\bf Author contributions: }

All authors contributed to this paper and/or carried out infrastructure work that made this analysis possible. GG led the work, performing the majority of the analysis and preparing the manuscript. GC conducted the full shear-ratio analysis and authored the corresponding section of the paper. AW provided support with validation tests, including those related to the scale-dependent response and PSF systematics. JP contributed to overall guidance throughout the project, assisted in preparing the manuscript, and provided specific support on the modelling framework. JB, CS, and GZ contributed to manuscript preparation as collaboration internal reviewers. AAl provided guidance on the shear-ratio analysis implementation and interpretation. EL contributed to early-stage code development used in this analysis.
GG, BY, WdA, GB, AAm, RC, AC, JM, MM, DG, and MG contributed to the calibration of photometric redshifts. DA, CT, and MRB lead the development of image and N-body simulations and derived calibrations. KB, ISN, ADW, ACR, RAG, SM, FM, AR, and ESR contributed to catalog infrastructure to provide the gold dataset. MRM and NW contributed to the \maglimpp sample, masks, and systematics calibration. MG, SS, TS, MY contributed to the shear calibration of the \mdet sample. AF, DSC, JB, JmC, contributed to developing the modelling components, analysis infrastructure, and validation tools. MJ developed the \textsc{Treecorr} code used in this analysis. AAm, SA, MRB, CC, MC, AP, JP, and Troxel contributed as coordinators of various groups and analyses related to this work. 
The remaining authors have made contributions to this paper that include, but are not limited to, the construction of DECam and other aspects of collecting the data; data processing and calibration; catalog creation; developing broadly used methods, codes, and simulations; running the pipelines and validation tests; and promoting the science analysis.

{\bf Funding and Institutional Support:} 

Funding for the DES Projects has been provided by the U.S. Department of Energy, the U.S. National Science Foundation, the Ministry of Science and Education of Spain, 
the Science and Technology Facilities Council of the United Kingdom, the Higher Education Funding Council for England, the National Center for Supercomputing 
Applications at the University of Illinois at Urbana-Champaign, the Kavli Institute of Cosmological Physics at the University of Chicago, 
the Center for Cosmology and Astro-Particle Physics at the Ohio State University,
the Mitchell Institute for Fundamental Physics and Astronomy at Texas A\&M University, Financiadora de Estudos e Projetos, 
Funda{\c c}{\~a}o Carlos Chagas Filho de Amparo {\`a} Pesquisa do Estado do Rio de Janeiro, Conselho Nacional de Desenvolvimento Cient{\'i}fico e Tecnol{\'o}gico and 
the Minist{\'e}rio da Ci{\^e}ncia, Tecnologia e Inova{\c c}{\~a}o, the Deutsche Forschungsgemeinschaft and the Collaborating Institutions in the Dark Energy Survey. 

The Collaborating Institutions are Argonne National Laboratory, the University of California at Santa Cruz, the University of Cambridge, Centro de Investigaciones Energ{\'e}ticas, 
Medioambientales y Tecnol{\'o}gicas-Madrid, the University of Chicago, University College London, the DES-Brazil Consortium, the University of Edinburgh, 
the Eidgen{\"o}ssische Technische Hochschule (ETH) Z{\"u}rich, 
Fermi National Accelerator Laboratory, the University of Illinois at Urbana-Champaign, the Institut de Ci{\`e}ncies de l'Espai (IEEC/CSIC), 
the Institut de F{\'i}sica d'Altes Energies, Lawrence Berkeley National Laboratory, the Ludwig-Maximilians Universit{\"a}t M{\"u}nchen and the associated Excellence Cluster Universe, 
the University of Michigan, NSF NOIRLab, the University of Nottingham, The Ohio State University, the University of Pennsylvania, the University of Portsmouth, 
SLAC National Accelerator Laboratory, Stanford University, the University of Sussex, Texas A\&M University, and the OzDES Membership Consortium.

Based in part on observations at NSF Cerro Tololo Inter-American Observatory at NSF NOIRLab (NOIRLab Prop. ID 2012B-0001; PI: J. Frieman), which is managed by the Association of Universities for Research in Astronomy (AURA) under a cooperative agreement with the National Science Foundation.

The DES data management system is supported by the National Science Foundation under Grant Numbers AST-1138766 and AST-1536171.
Data access is enabled by Jetstream2 and OSN at Indiana University through allocation PHY240006: Dark Energy Survey from the Advanced Cyberinfrastructure Coordination Ecosystem: Services and Support (ACCESS) program, which is supported by U.S. National Science Foundation grants 2138259, 2138286, 2138307, 2137603, and 2138296.
The DES participants from Spanish institutions are partially supported by MICINN under grants PID2021-123012, PID2021-128989 PID2022-141079, SEV-2016-0588, CEX2020-001058-M and CEX2020-001007-S, some of which include ERDF funds from the European Union. IFAE is partially funded by the CERCA program of the Generalitat de Catalunya.

We  acknowledge support from the Brazilian Instituto Nacional de Ci\^encia
e Tecnologia (INCT) do e-Universo (CNPq grant 465376/2014-2).

This document was prepared by the DES Collaboration using the resources of the Fermi National Accelerator Laboratory (Fermilab), a U.S. Department of Energy, Office of Science, Office of High Energy Physics HEP User Facility. Fermilab is managed by Fermi Forward Discovery Group, LLC, acting under Contract No. 89243024CSC000002.

% Do not erase.
GG acknowledges support from the grant JDC2023-052892-I, funded by MCIU/AEI/10.13039/501100011033 and by the ESF+.

GC and AA acknowledge support from MICIU/AEI grant PID2023-153229NA and a grant by LaCaixa Foundation (ID 100010434) code LCF/BQ/PI23/11970028.

\section{Data Availability}
The DES Y6 data products used in this work are publicly available at \citep{y6-data}. As cosmology likelihood sampling software we use \texttt{cosmosis}, available at \citep{cosmosis-github}.



\bibliography{main, y6kp}

\appendix
\section{Tangential shear around random points}\label{app:random_points}

\begin{figure*}
    \centering
    \includegraphics[width=\textwidth]{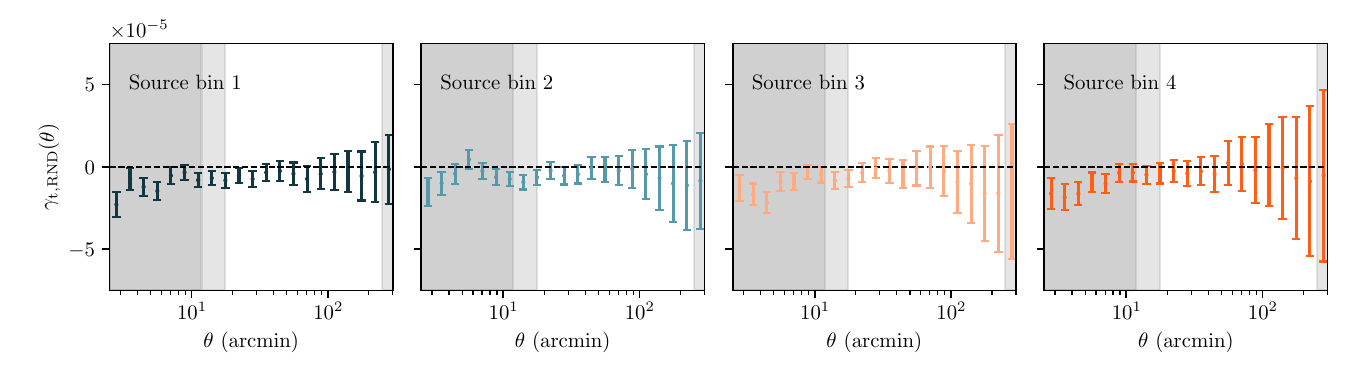}
    \caption{Tangential shear measured around random points, for each source tomographic bin. The error bars are obtained from the Jackknife covariance. Gray shaded regions mark the angular scales removed from the analysis: darker gray for the linear-bias scale cuts and lighter gray for the non-linear-bias cuts. Lens bin 2 is fully shaded because it is excluded from the fiducial data vector. }
    \label{fig:random_points}
\end{figure*}

In Section \ref{sec:measurement}, we described how the tangential-shear estimator is affected by additive shear systematics arising from residual mean shear, PSF-related terms, and selection anisotropies. In Figure \ref{fig:random_points}, we show the tangential shear measured around random points for each source tomographic bin, using the random catalog associated with the first lens bin. A slight residual signal is visible at small angular scales. This behavior is expected: the shear catalog contains small spatially coherent additive components, and the random-point measurement simply captures their tangential projection when averaged over the survey mask. The amplitude of this residual is well below the statistical uncertainty of the galaxy-galaxy lensing measurement on the scales used for cosmology, and its presence is the reason why random-point subtraction is included in the estimator.

\begin{figure*}
    \centering
    \includegraphics[width=\textwidth]{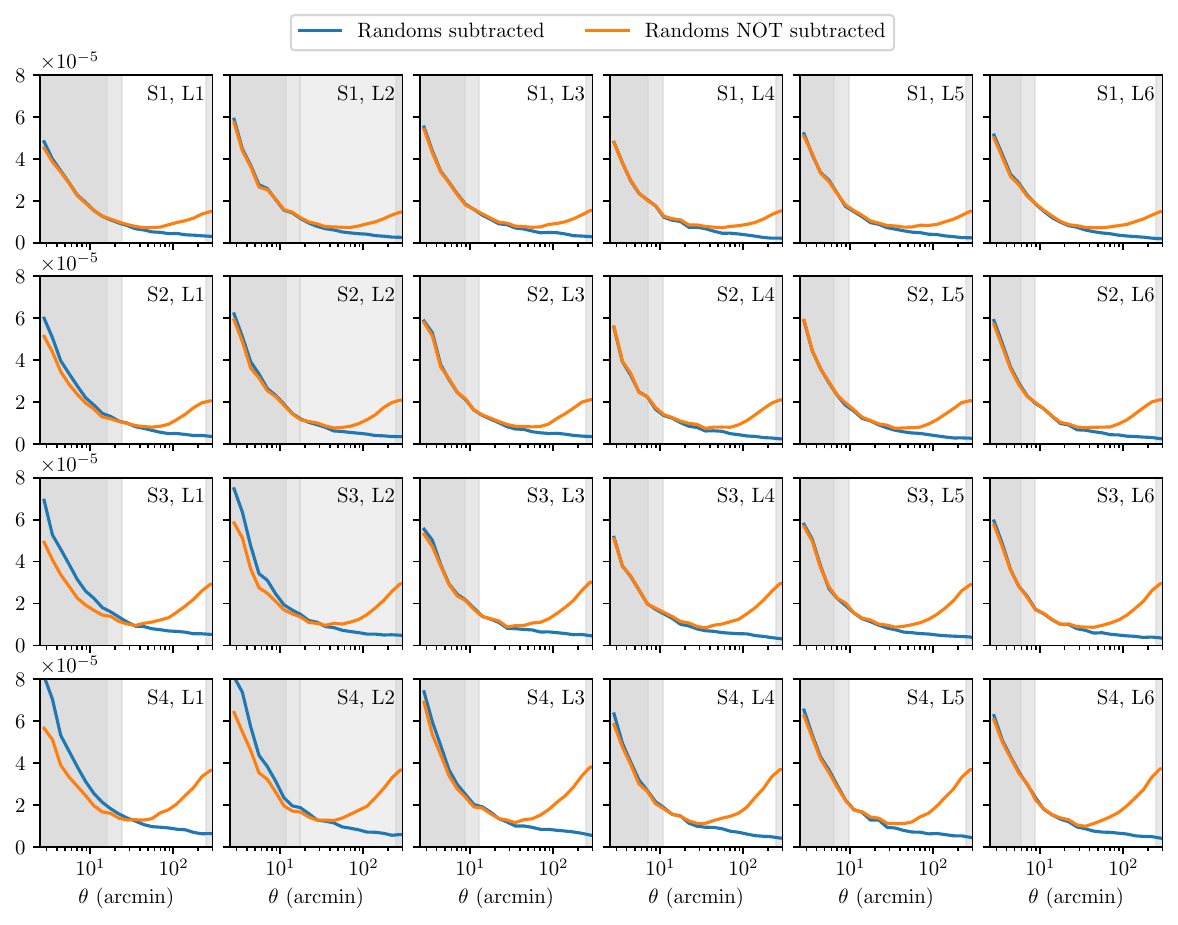}
    \caption{Jackknife covariance comparison when subtracting or not the tangential shear around random points from the baseline estimator. Gray shaded regions mark the angular scales removed from the analysis: darker gray for the linear-bias scale cuts and lighter gray for the non-linear-bias cuts. Lens bin 2 is fully shaded because it is excluded from the fiducial data vector. }
    \label{fig:cov_randoms}
\end{figure*}

In Figure \ref{fig:cov_randoms}, we show the jackknife covariance for each lens-source bin pair and the effect of subtracting the random-point contribution. As discussed in \cite{Singh2017}, the subtraction reduces large-scale covariance by removing additive shear terms that are coherent across the footprint, leading to a more stable and less noisy estimator of the galaxy-galaxy lensing signal. On the large angular scales, the subtraction substantially reduces the covariance, as expected.

\section{Intrinsic Alignment parameters posteriors}\label{app:ggl_ia}

\begin{figure}
    \centering
    \includegraphics[width=0.47\textwidth]{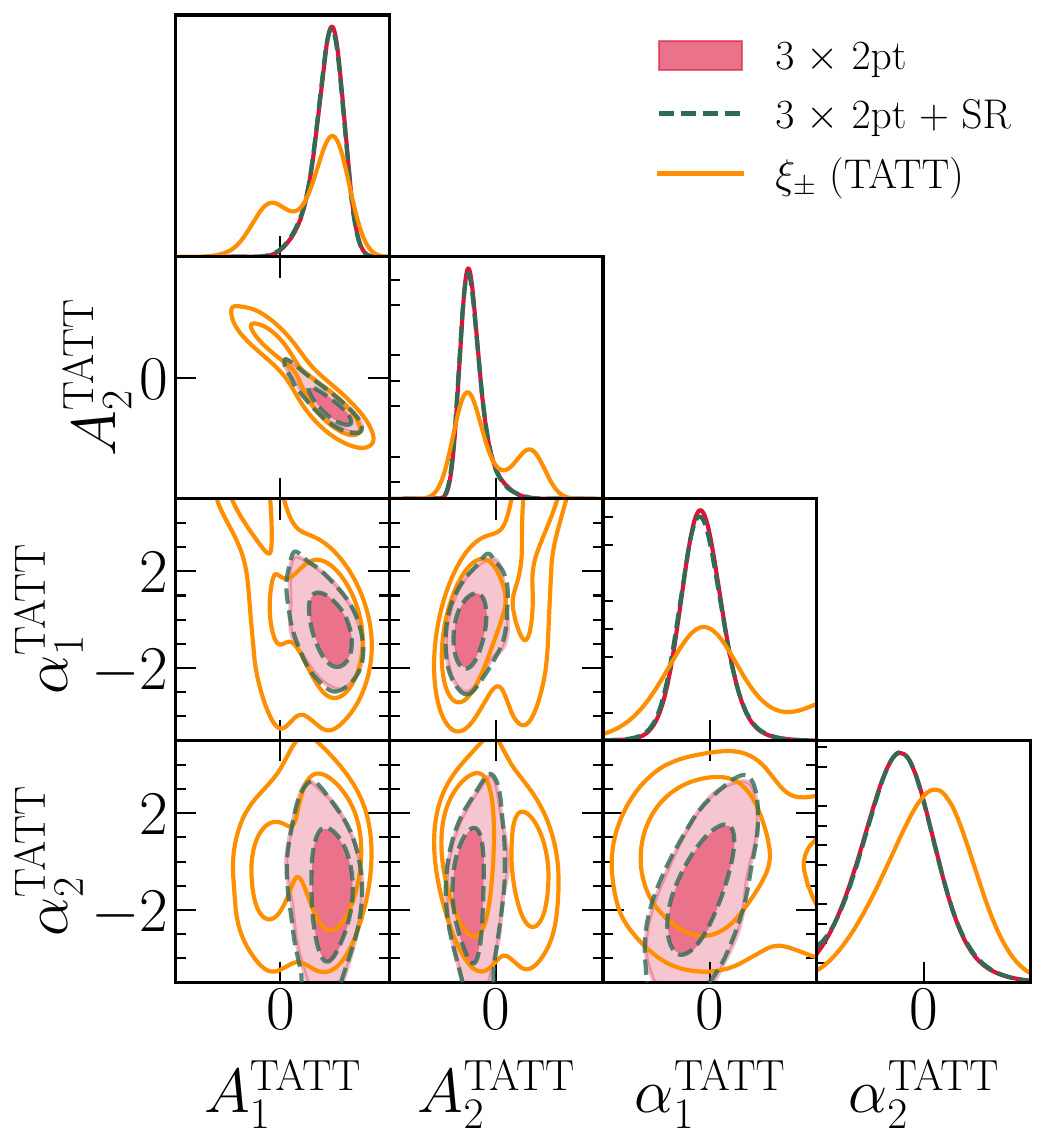}
    \caption{Constraints on the four TATT intrinsic-alignment parameters derived from different data combinations. Orange contours show constraints from cosmic shear $\xi_\pm$ alone, while red contours correspond to the full DES Y6 $3\times2\mathrm{pt}$ analysis. Dashed green contours indicate the $3\times2\mathrm{pt}$ analysis augmented with shear-ratio (SR) information. The posterior distributions for $(A^{\mathrm{TATT}}_1, A^{\mathrm{TATT}}_2, \alpha^{\mathrm{TATT}}_1, \alpha^{\mathrm{TATT}}_2)$ illustrate how the inclusion of galaxy--galaxy lensing significantly tightens constraints by directly probing correlations between source shapes and the lens density field.}
    \label{fig:ia_posteriors}
\end{figure}

Intrinsic alignments enter the weak-lensing analysis through correlations between galaxy shapes and the underlying matter field. In the TATT model, these effects are encoded in a set of amplitude and scale-dependence parameters $(A^{\mathrm{TATT}}_1, A^{\mathrm{TATT}}_2, \alpha^{\mathrm{TATT}}_1, \alpha^{\mathrm{TATT}}_2)$, which describe the response of galaxy shapes to the large-scale tidal field at first and second order. While cosmic shear provides direct sensitivity to a mixture of II and GI contributions, the GGL signal supplies a complementary source of information that is crucial for constraining the full IA model.

Galaxy--galaxy lensing probes the tangential shear of background sources around lens galaxies, which serve as a direct tracer of the density field. IA affects this observable in two ways. First, correlations between intrinsically aligned source shapes and the positions of lens galaxies generate an additional GI-type term in the predicted signal. Second, IA modifies the redshift and scale weighting of the shear field in a way that differs from the purely gravitational contribution. Because GGL isolates the alignment of galaxy shapes with the lens density field---rather than with other source galaxies as in cosmic shear---it provides sensitivity to distinct linear combinations of the TATT parameters. This shifts and contracts the allowed parameter region when GGL is included in the analysis.

In DES Y6, the high signal-to-noise ratio of the GGL measurement and the wide range of lens--source redshift separations significantly strengthen our ability to break degeneracies between IA parameters and cosmological parameters such as $\sigma_8$ and $\Omega_m$. Each lens--source bin pair samples a different geometric configuration, which in turn yields differential sensitivity to the scale-dependence parameters $\alpha^{\mathrm{TATT}}_1$ and $\alpha^{\mathrm{TATT}}_2$. As a result, adding GGL forces the IA model to simultaneously reproduce (i) the shear--shear correlations present in cosmic shear and (ii) the shape--density correlations present in GGL. This joint requirement eliminates flat directions in the IA parameter space that are present in cosmic shear alone, particularly in directions linked to the nonlinear evolution of the matter power spectrum.

The improvement is clearly visible in Fig.~\ref{fig:ia_posteriors}, where the posterior distributions from cosmic shear alone (orange) exhibit substantially wider degeneracy directions compared to the full $3\times2\mathrm{pt}$ constraints (pink). The inclusion of GGL noticeably reduces the allowed volume in $(A^{\mathrm{TATT}}_1, A^{\mathrm{TATT}}_2)$ and tightens the constraints on $(\alpha^{\mathrm{TATT}}_1, \alpha^{\mathrm{TATT}}_2)$ by linking IA-induced distortions in the shear field to observable fluctuations in the lens density field.

Overall, GGL plays a central role in stabilising and tightening the IA constraints in DES Y6. Even when cosmic shear alone permits broad or highly degenerate regions in TATT parameter space, the addition of GGL forces IA parameters to match the observed shape--density correlations, thereby producing a substantially more robust inference of the IA sector.

\section{Code Comparison}\label{app:code_comparison}

To ensure the robustness and reproducibility of our galaxy-galaxy lensing measurements, we performed a detailed comparison between the pipeline used in this work and an independent fiducial pipeline that performs the standard measurement of all three two-point correlation functions ($\gamma_t$, $\xi_\pm$, and $w(\theta)$). The fiducial pipeline is designed to reproduce the baseline measurement without the additional validation and systematic tests described in this paper.

TreeCorr accelerates correlation calculations by allowing small tolerances in the assignment of pairs to separation bins and in the angular projection used for shear quantities, controlled by the parameters \texttt{bin\_slop} and \texttt{angle\_slop}. In our development tests we adopted small values (\texttt{bin\_slop}=0.01, \texttt{angle\_slop}=0.05), which significantly reduce the computational cost while introducing only sub-percent level shifts in the accumulated pairs. We verified this by reducing the slop parameters by a factor of two and confirming that the resulting correlation functions remained unchanged within numerical precision. For this code-comparison, we set both parameters to zero, which forces TreeCorr to place each pair into its exact separation bin and to use the exact pair orientation for shear projections. This ensures maximal numerical accuracy for all final results.

% The comparison was carried out in two stages. First, we performed the test using the blinded data vector to verify the consistency of the measurements without knowledge of the true signal. After unblinding, we repeated the test with the true data vector to confirm that the agreement persisted. In both cases, we adopted the most accurate \textsc{TreeCorr} configuration, with \texttt{bin\_slop} and \texttt{angle\_slop} set to zero, to eliminate any numerical variability associated with approximate binning.

%

\begin{figure*}
    \centering
    \includegraphics[width=\linewidth]{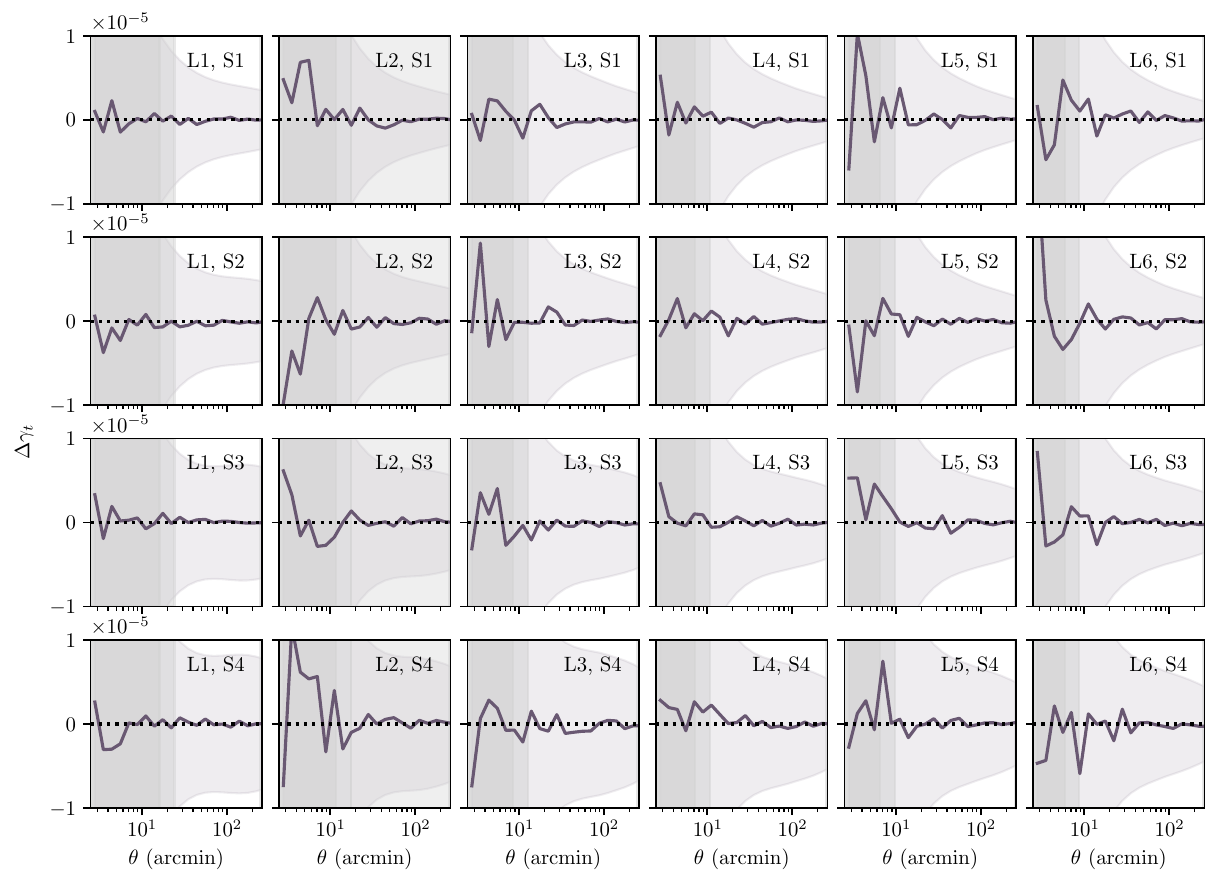}
    \caption{Residual between the fiducial pipeline and the pipeline used in this work. The transparent purple filled area represents the uncertainty estimated from the theory covariance. Gray shaded regions mark the angular scales removed from the analysis: darker gray for the linear-bias scale cuts and lighter gray for the non-linear-bias cuts. Lens bin 2 is fully shaded because it is excluded from the fiducial data vector. Note that this figure does not illustrate the point-mass marginalization, which is applied during likelihood evaluation through the inverse covariance matrix and therefore cannot be represented in terms of the covariance elements shown here.}
    \label{fig:codecomp}
\end{figure*}

The results of this comparison, shown in Figure~\ref{fig:codecomp}, demonstrate excellent agreement between the two pipelines, with differences consistent with numerical precision effects due to running on different machines. 
The fiducial pipeline will be publicly available at the \url{https://www.darkenergysurvey.org/des-y6-cosmology-results-papers/}.

\label{lastpage}
\end{document}